\documentclass[journal]{IEEEtran}

\usepackage{cite}

\usepackage[pdftex]{graphicx}
\makeatletter
\let\MYcaption\@makecaption
\makeatother

\usepackage[font=footnotesize]{subcaption}

\makeatletter
\let\@makecaption\MYcaption
\makeatother
\usepackage{amsmath,amssymb,amsfonts, bm}
\usepackage{amsthm}
\usepackage{chngcntr}
\newtheorem{thm}{Theorem}
\newtheorem{cor}{Corollary}

\newtheorem{asmp}{Assumption}
\newtheorem{cond}{Condition}
\newtheorem{lem}{Lemma}
\theoremstyle{definition}
\newtheorem{rem}{Remark}
\newtheorem{ex}{Example}
\usepackage[ruled,norelsize]{algorithm2e}
\SetKwBlock{Repeat}{repeat}{}
\usepackage{array}
\usepackage{bbm}
\usepackage{comment}
\usepackage{hhline}
\usepackage[flushleft]{threeparttable}
\usepackage{nccmath}
\usepackage{multicol}
\usepackage{makecell}
\usepackage{tabularx}
\usepackage[hidelinks]{hyperref}
\usepackage{xcolor}

\newcommand{\colspace}[1]{\mathcal{{C}}(#1)}
\newcommand{\mc}[1]{\mathcal{#1}}

\newcommand{\norm}[1]{\left\lVert#1\right\rVert}

\makeatletter
\let\save@mathaccent\mathaccent
\newcommand*\if@single[3]{%
  \setbox0\hbox{${\mathaccent"0362{#1}}^H$}%
  \setbox2\hbox{${\mathaccent"0362{\kern0pt#1}}^H$}%
  \ifdim\ht0=\ht2 #3\else #2\fi
  }
\newcommand*\rel@kern[1]{\kern#1\dimexpr\macc@kerna}
\newcommand*\widebar[1]{\@ifnextchar^{{\wide@bar{#1}{0}}}{\wide@bar{#1}{1}}}
\newcommand*\wide@bar[2]{\if@single{#1}{\wide@bar@{#1}{#2}{1}}{\wide@bar@{#1}{#2}{2}}}
\newcommand*\wide@bar@[3]{%
  \begingroup
  \def\mathaccent##1##2{%
    \let\mathaccent\save@mathaccent
    \if#32 \let\macc@nucleus\first@char \fi
    \setbox\z@\hbox{$\macc@style{\macc@nucleus}_{}$}%
    \setbox\tw@\hbox{$\macc@style{\macc@nucleus}{}_{}$}%
    \dimen@\wd\tw@
    \advance\dimen@-\wd\z@
    \divide\dimen@ 3
    \@tempdima\wd\tw@
    \advance\@tempdima-\scriptspace
    \divide\@tempdima 10
    \advance\dimen@-\@tempdima
    \ifdim\dimen@>\z@ \dimen@0pt\fi
    \rel@kern{0.6}\kern-\dimen@
    \if#31
      \overline{\rel@kern{-0.6}\kern\dimen@\macc@nucleus\rel@kern{0.4}\kern\dimen@}%
      \advance\dimen@0.4\dimexpr\macc@kerna
      \let\final@kern#2%
      \ifdim\dimen@<\z@ \let\final@kern1\fi
      \if\final@kern1 \kern-\dimen@\fi
    \else
      \overline{\rel@kern{-0.6}\kern\dimen@#1}%
    \fi
  }%
  \macc@depth\@ne
  \let\math@bgroup\@empty \let\math@egroup\macc@set@skewchar
  \mathsurround\z@ \frozen@everymath{\mathgroup\macc@group\relax}%
  \macc@set@skewchar\relax
  \let\mathaccentV\macc@nested@a
  \if#31
    \macc@nested@a\relax111{#1}%
  \else
    \def\gobble@till@marker##1\endmarker{}%
    \futurelet\first@char\gobble@till@marker#1\endmarker
    \ifcat\noexpand\first@char A\else
      \def\first@char{}%
    \fi
    \macc@nested@a\relax111{\first@char}%
  \fi
  \endgroup
}
\makeatother

\newcommand{\wip}[1]{{\color{blue}#1}}

\newcolumntype{Y}{>{\centering\arraybackslash}X}
\newcolumntype{b}{>{\hsize=1.35\hsize}Y}
\newcolumntype{z}{>{\hsize=1.2\hsize}Y}
\newcolumntype{s}{>{\hsize=.45\hsize}Y}

\makeatletter
\newcommand*{\inlineequation}[2][]{%
  \begingroup
    \refstepcounter{equation}%
    \ifx\\#1\\%
    \else
      \label{#1}%
    \fi
    \relpenalty=10000 %
    \binoppenalty=10000 %
    \ensuremath{%
      #2%
    }%
    ~\@eqnnum
  \endgroup
}
\makeatother
\makeatletter
\newcommand{\removelatexerror}{\let\@latex@error\@gobble}
\makeatother

\newcounter{parentasmp}
\makeatletter 
\newenvironment{subasmp}[1]{%
  \counterwithin*{asmp}{parentasmp}
  \def\subasmpcounter{#1}%
  \refstepcounter{#1}%
  \protected@edef\theparentasmp{\csname the#1\endcsname}%
  \setcounter{parentasmp}{\value{#1}}%
  \setcounter{#1}{0}%
  \expandafter\def\csname the#1\endcsname{\theparentasmp\alph{#1}}%
  \ignorespaces
}{%
  \setcounter{\subasmpcounter}{\value{parentasmp}}%
  \counterwithout*{asmp}{parentasmp} 
  \ignorespacesafterend
}
\makeatother

\newcounter{parentcond}
\makeatletter 
\newenvironment{subcond}[1]{%
  \counterwithin*{cond}{parentcond}
  \def\subcondcounter{#1}%
  \refstepcounter{#1}%
  \protected@edef\theparentcond{\csname the#1\endcsname}%
  \setcounter{parentcond}{\value{#1}}%
  \setcounter{#1}{0}%
  \expandafter\def\csname the#1\endcsname{\theparentcond\alph{#1}}%
  \ignorespaces
}{%
  \setcounter{\subcondcounter}{\value{parentcond}}%
  \counterwithout*{cond}{parentcond} 
  \ignorespacesafterend
}
\makeatother

\SetKwFor{At}{at}{do}{end}

\begin{document}

\setlength{\abovedisplayskip}{4pt}
\setlength{\belowdisplayskip}{4pt}
\setlength{\textfloatsep}{5pt plus 1.0pt minus 2.0pt}

\title{A Unified Algorithmic Framework for Distributed Adaptive Signal and Feature Fusion Problems \\\LARGE--- Part II: Convergence Properties}

\author{Cem Ates~Musluoglu, Charles Hovine,
        and~Alexander~Bertrand,~\IEEEmembership{Senior~Member,~IEEE}
\thanks{Copyright \copyright 2023 IEEE. Personal use of this material is permitted. Permission from IEEE must be obtained for all other uses, in any current or future media, including reprinting/republishing this material for advertising or promotional purposes, creating new collective works, for resale or redistribution to servers or lists, or reuse of any copyrighted component of this work in other works.}
\thanks{This project has received funding from the European Research Council (ERC) under the European Union's Horizon 2020 research and innovation programme (grant agreement No 802895). The authors also acknowledge the financial support of the FWO (Research Foundation Flanders) for project G081722N, and the Flemish Government (AI Research Program).}
\thanks{C.A. Musluoglu, C. Hovine and A. Bertrand are with KU Leuven, Department of Electrical Engineering (ESAT), Stadius Center for Dynamical Systems, Signal Processing and Data Analytics, Kasteelpark Arenberg 10, box 2446, 3001 Leuven, Belgium and with Leuven.AI - KU Leuven institute for AI. e-mail: cemates.musluoglu, charles.hovine, alexander.bertrand @esat.kuleuven.be}
\thanks{A companion paper submitted together with this paper is provided in \cite{musluoglu2022unifiedp1}. This paper has supplementary downloadable material \cite{musluoglu2022unifiedp2supp} available at \url{http://ieeexplore.ieee.org}, provided by the authors. The material includes the steps we follow to prove Lemma \ref{lem:rank_H}. This material is 397 KB in size.}
\thanks{Digital Object Identifier: 10.1109/TSP.2023.3275273}
}

\maketitle

\begin{abstract}
	This paper studies the convergence conditions and properties of the distributed adaptive signal fusion (DASF) algorithm, the framework itself having been introduced in a `Part I' companion paper. The DASF algorithm can be used to solve linear signal and feature fusion optimization problems in a distributed fashion, and is in particular well-suited for solving spatial filtering optimization problems encountered in wireless sensor networks. The convergence conditions and results are provided along with rigorous proofs and analyses, as well as various example problems to which they apply. Additionally, we describe procedures that can be added to the DASF algorithm to ensure convergence in specific cases where some of the technical convergence conditions are not satisfied.
	\end{abstract}
	
\section{Introduction}
\IEEEPARstart{T}{he} Distributed Adaptive Signal Fusion (DASF) algorithm introduced in \cite{musluoglu2022unifiedp1} can be used to solve a wide range of spatial filtering and signal fusion problems in a distributed fashion, e.g., within a wireless sensor network (WSN). Examples of such problems include minimum mean square error estimation, discriminant analysis based on generalized eigenvalue decomposition \cite{liu1992generalized}, canonical correlation analysis \cite{borga1998learning,hardoon2004canonical}, minimum variance beamforming \cite{van1988beamforming}, etc. The DASF algorithm is designed to cope with the typical bandwidth or energy limitations of WSNs.

A typical spatial filtering or signal fusion problem in a WSN involves optimizing a cost function depending on the sensor data collected by each node in the network. Contrarily to a centralized procedure requiring the sensor data of each node to be aggregated at a fusion center, the DASF algorithm requires the nodes to share only compressed data between each other. This data is then used to locally build a compressed version of the global optimization problem within a node at every iteration. As a result, any solver for the global (centralized) problem can be used to solve the local problems at each node within the DASF iterations. 

In this paper, we provide a set of sufficient conditions for convergence and optimality of the DASF algorithm, based on which we can show that the DASF algorithm converges to the centralized solution of the problem despite the compression, as if all the raw sensor data were centrally available. The technical conditions required for convergence are akin to the well-known linear independence constraint qualifications in the optimization literature, which in the case of DASF lead to an upper bound on the number of constraints the global (centralized) problem is allowed to have. Furthermore, since the local problems in each node are compressed versions of the global problem, we assume that the local problems satisfy the same assumptions as the global problem as outlined in \cite{musluoglu2022unifiedp1}, which is generally the case as they are directly inherited. Finally, we impose a condition on the finiteness of the number of possible solutions that are achievable by the solver used to solve the local problems in each node. We will see that these conditions are often satisfied for spatial filtering and signal fusion problems in practical scenarios. Furthermore, we provide several examples and illustrations on how the convergence conditions can be checked either a priori or during operation of the algorithm. We will also show how the insights obtained from the convergence analysis can be used to design new strategies to enforce convergence in cases where a violation of the convergence conditions is detected.

The outline of the paper is as follows. After a short review of the DASF framework in Section \ref{sec:dsfo_review}, we study the convergence and optimality guarantees of the DASF algorithm in Section \ref{sec:conv_opt}. In particular, we show that under some technical conditions, accumulation points of the sequence of points produced by the algorithm are also fixed points, and that fixed points are solutions of the centralized problem. Examples of typical spatial filtering and signal fusion problems, such as minimum mean square error or minimum variance beamforming, and how the convergence conditions apply to these cases are discussed in Section \ref{sec:examples}. Finally, in the contrived cases where some of the technical requirements are violated, we describe methods to still achieve convergence for the DASF algorithm in Section \ref{sec:extra}. Conclusions are drawn in Section \ref{sec:conclusion}.

\textbf{Notation:} Uppercase letters are used to represent matrices and sets, the latter in calligraphic script, while scalars, scalar-valued functions and vectors are represented by lowercase letters, the latter in bold. We use the notation $\chi_q^i$ to refer to a certain mathematical object $\chi$ (such as a matrix, set, etc.) at node $q$ and iteration $i$. The notation $\left(\chi^i\right)_{i\in\mathcal{I}}$ refers to a sequence of elements $\chi^i$ over every index $i$ in the ordered index set $\mathcal{I}$. If it is clear from the context (often in the case where $i$ is over all natural numbers), we omit the index set $\mathcal{I}$ and simply write $\left(\chi^i\right)_i$. A similar notation $\{\chi^i\}_{i\in\mathcal{I}}$ is used for non-ordered sets. Additionally, $I_Q$ denotes the $Q\times Q$ identity matrix, $\mathbb{E}[\cdot]$ the expectation operator, $\text{tr}(\cdot)$ the trace operator, $\textit{BlkDiag}\left(\cdot\right)$ the operator that creates a block-diagonal matrix from its arguments and $|\cdot|$ the cardinality of a set.

\section{Review of the DASF Framework}\label{sec:dsfo_review}

In this section, we briefly restate the scope and operation of the DASF framework, which was extensively described in \cite{musluoglu2022unifiedp1}. We do this for completeness and with the purpose to re-introduce key equations and introduce some new ones that will be needed in the convergence analysis. The reader who is not yet familiar with the DASF framework is encouraged to read \cite{musluoglu2022unifiedp1} first, since the review in this section skips several useful insights and details. 

\subsection{Problem Description and Assumptions}\label{sec:prob_descript}

We consider a WSN consisting of a set of $K$ nodes $\mathcal{K}=\{1,\dots,K\}$ interconnected according to a network graph $\mathcal{G}$, which contains edges between nodes that are able to share data with each other. Each node $k$ collects observations of an $M_k-$channel sensor signal $\mathbf{y}_k$. We define the network-wide sensor signal $\mathbf{y}\in\mathbb{R}^{M}$ as
\begin{equation}\label{eq:y_part}
  \mathbf{y}(t)=[\mathbf{y}_1^T(t),\dots,\mathbf{y}_K^T(t)]^T,
\end{equation}
where $t$ denotes the time index and $M=\sum_{k\in\mathcal{K}}M_k$. $\mathbf{y}$ is assumed to be (short-term) stationary and ergodic, such that its statistical properties can be properly estimated given a sufficiently large number of samples at different time instances, e.g., $\{\mathbf{y}(\tau)\}_{\tau=0}^{N-1}$, where $N$ denotes the number of time samples. Our objective is to find a linear filter $X\in\mathbb{R}^{M\times Q}$ that optimizes in some sense the output signal of the linear signal fusion $X^T\mathbf{y}$. More specifically, we consider problems of the following form: 
\begin{equation}\label{eq:prob_g}
  \begin{aligned}
  \mathbb{P}:\underset{X\in\mathbb{R}^{M\times Q}}{\text{minimize } } \quad & \varphi\left(X^T\mathbf{y}(t),X^TB\right)\\
  \textrm{subject to} \quad & \eta_j\left(X^T\mathbf{y}(t),X^TB\right)\leq 0\;\textrm{ $\forall j\in\mathcal{J}_I$,}\\
    & \eta_j\left(X^T\mathbf{y}(t),X^TB\right)=0\;\textrm{ $\forall j\in\mathcal{J}_E$,}
  \end{aligned}
\end{equation}
where $\varphi$ and $\eta_j$'s are differentiable real- and scalar-valued functions. We refer to such problems as \textit{(distributed) signal fusion optimization} ((D)SFO) problems. In Problem (\ref{eq:prob_g}), $\varphi$ denotes the objective function and the $\eta_j$'s denote the constraint functions, where the indices $j\in\mathcal{J}_I$ and $j\in\mathcal{J}_E$ correspond to inequality and equality constraints respectively. We denote the joint index set $\mathcal{J}_I\cup \mathcal{J}_E=\mathcal{J}$, such that the total number of constraints is $J=|\mathcal{J}|$. 

The objective and constraint functions implicitly contain an operator that maps the stochastic variable $\mathbf{y}$ into a deterministic and time-independent quantity, such as an expectation operator, which in practice is typically replaced with an estimated quantity based on temporal averaging of $N$ samples of $\mathbf{y}$.

$B$ is an additional deterministic parameter of the problem. Similarly to (\ref{eq:y_part}), it can be partitioned as
\begin{equation}\label{eq:B_part}
  B=[B_1^T,\dots,B_K^T]^T\in\mathbb{R}^{M\times L}
\end{equation}
where $L$ is some problem-dependent constant. It is assumed that each node $k$ has a local access to its corresponding block $B_k$ (i.e. they can be used in computations without being first requested from another node). In contrast to the signal $\mathbf{y}$, the matrix $B$ is deterministic and independent of time. As noted in \cite{musluoglu2022unifiedp1}, Problem (\ref{eq:prob_g}) can be generalized to more than one variable $X$, stochastic signal $\mathbf{y}$ and deterministic term $B$. Note that such a generalization covers deterministic quadratic terms in the form $X^TAX$, since two linear terms $X^TB^{(1)}$ and $X^TB^{(2)}$ can be combined into $(X^TB^{(1)})\cdot (X^TB^{(2)})^T$, with $A=B^{(1)}B^{(2)T}$.

As the actual optimization variable is $X$, and by removing the time-dependence of the problem by stationarity of the random signal $\mathbf{y}$, we define the functions $f$ and $h_j$, $j\in\mathcal{J}$, which express the objective and constraints as a function of $X$ only:
\begin{align}
  f(X)&\triangleq\varphi\left(X^T\mathbf{y}(t),X^TB\right),\label{eq:f_simplify}\\
  h_j(X)&\triangleq\eta_j\left(X^T\mathbf{y}(t),X^TB\right),\;\forall j\in\mathcal{J},\label{eq:hj_simplify}
\end{align}
which are assumed to be continuously differentiable with respect to the variable $X$ over their respective domains. This allows us to write (\ref{eq:prob_g}) compactly as
\begin{equation}\label{eq:prob_g_compact}
  \begin{aligned}
  \mathbb{P}:\underset{X\in\mathbb{R}^{M\times Q}}{\text{minimize } } \quad & f(X)\\
  \textrm{subject to} \quad & h_j(X)\leq 0\;\textrm{ $\forall j\in\mathcal{J}_I$,}\\
    & h_j(X)=0\;\textrm{ $\forall j\in\mathcal{J}_E$.}
  \end{aligned}
\end{equation}
Furthermore, we denote the constraint set of (\ref{eq:prob_g}) or (\ref{eq:prob_g_compact}) as $\mathcal{S}$, the complete solution set as $\mathcal{X}^*$ and a single solution as $X^*$, i.e., $X^*\in\mathcal{X}^*$. 

In order to guarantee the theoretical convergence of the DASF algorithm, we restrict its application to problems satisfying the following three general assumptions\footnote{Throughout this text, if assumptions or conditions are labeled as ``$\textbf{Xa}$'', it implies that this assumption/condition can be replaced with a different assumption/condition ``$\textbf{Xb}$''. When we only mention the label ``$\textbf{X}$'', we refer to either of the two.} (in Section \ref{sec:examples}, we explain how these assumptions can be checked in practical examples):
\begin{asmp}\label{asmp:well_posed}
        The targeted instance of Problem (\ref{eq:prob_g}) or (\ref{eq:prob_g_compact}) is well-posed, in the sense that the solution set is not empty and varies continuously with a change in the parameters of the problem.
\end{asmp}
\noindent The notion of (generalized Hadamard) well-posedness we require is
based on \cite{hadamard1902problemes,zhou2005hadamard}. The main difference is
that we require the map from the space of inputs of the problem to the solution
space to be continuous instead of upper semicontinuous, which is required for
the convergence proof. We formally define the continuity of this map in Section
\ref{sec:conv_opt}. Even though this condition might seem restrictive, it applies
to many practical instances of the problems of interest (see Section \ref{sec:examples}).
\begin{asmp}\label{asmp:kkt}
  The linear independence constraint qualifications (LICQ)
  \cite{peterson1973review} hold at the solutions of Problem (\ref{eq:prob_g}) or (\ref{eq:prob_g_compact}), i.e., the solutions satisfy the Karush-Kuhn-Tucker (KKT) conditions.
\end{asmp}

\noindent If $X^*$ is a solution of Problem (\ref{eq:prob_g}) or (\ref{eq:prob_g_compact}), Assumption \ref{asmp:kkt} implies that, for $j\in\mathcal{J}^*$, the gradients $\nabla_X h_j(X^*)$ are linearly independent\footnote{A set of matrices $\{A_j\}_{j\in\mathcal{J}}$ is linearly independent when $\sum_{j\in\mathcal{J}}\alpha_jA_j=0$ is satisfied if and only if $\alpha_j=0$, $\forall j\in\mathcal{J}$, or equivalently, when $\{\text{vec}(A_j)\}_{j\in\mathcal{J}}$ is a set of linearly independent vectors, where $\text{vec}(\cdot)$ is the vectorization operator.}, where $\mathcal{J}^*\subseteq \mathcal{J}$ is the set of all indices $j$ satisfying $h_j(X^*)=0$. If the problem is unconstrained, we have $\nabla_X f(X^*)=0$.

\begin{subasmp}{asmp}\label{asmp:compactness}
\begin{asmp}\label{asmp:compact_sublevel}
  $f$ has compact sublevel sets in $\mathcal{S}$, i.e., for all $m\in\mathbb{R}$, $\{X\in\mathcal{S}\;|\;f(X)\leq m\}$ is compact, i.e., closed and bounded.
\end{asmp}
It is noted here that this assumption can be further relaxed to an alternative Assumption \ref{asmp:compact_X0} presented below. This relaxed version only requires that the DASF algorithm's initialization point is in a compact sublevel set, in which case not all sublevel sets of $f$ in $\mathcal{S}$ should be compact. 

\begin{asmp}\label{asmp:compact_X0}
    The sublevel set of $f$ $\{X\in\mathcal{S}\;|\;f(X)\leq f(X^0)\}$ corresponding to $X^0$ is compact.
\end{asmp}
\end{subasmp}
\renewcommand*{\theHasmp}{\theasmp}
As will be shown in Section \ref{sec:obj_conv}, Assumption \ref{asmp:compact_sublevel} or \ref{asmp:compact_X0} is needed to ensure that the elements of the sequence $(X^i)_i$ obtained from the DASF algorithm lie in a compact set, which is required to show convergence. 

In the remaining of this paper, problems $\mathbb{P}$ which can be written as (\ref{eq:prob_g}) or (\ref{eq:prob_g_compact}) will be satisfying the assumptions above, i.e., we will not repeat these assumptions in any of the convergence theorems. 
As discussed in \cite{musluoglu2022unifiedp1}, we also implicitly assume that there exists a centralized solver able to solve the targeted problem instance $\mathbb{P}$, which will be used by the DASF algorithm to solve local per-node compressed versions of the centralized problem $\mathbb{P}$.

\subsection{The DASF Algorithm}
In order to solve Problem (\ref{eq:prob_g}) in a distributed setting, we consider a partitioning of the global variable $X$ into local variables:
\begin{equation}\label{eq:X_part}
  X=[X_1^T,\dots,X_K^T]^T,
\end{equation}
where each local variable $X_k\in\mathbb{R}^{M_k\times Q}$ is assigned to a single node $k$. At any given iteration $i$, the nodes $k$ all have an estimation $X_k^i$ of their local variable $X_k$. At each iteration $i$, some node $q\in\mathcal{K}$ is selected to be the ``updating node'', and will solve a compressed version of Problem (\ref{eq:prob_g}), which will be defined later in this section. We note that we choose a different updating node at each iteration. As the network graph $\mathcal{G}$ can contain loops, we prune the network into a tree, which we denote as $\mathcal{T}^i(\mathcal{G},q)$, such that there is a unique path from any node $k\neq q$ to the updating node $q$ (as explained in \cite{musluoglu2022unifiedp1}, the tree $\mathcal{T}^i(\mathcal{G},q)$ should preserve all the neighbors of node $q$ to maximize the degrees of freedom in the updating process).

\begin{figure}[t]
  \includegraphics[width=0.48\textwidth]{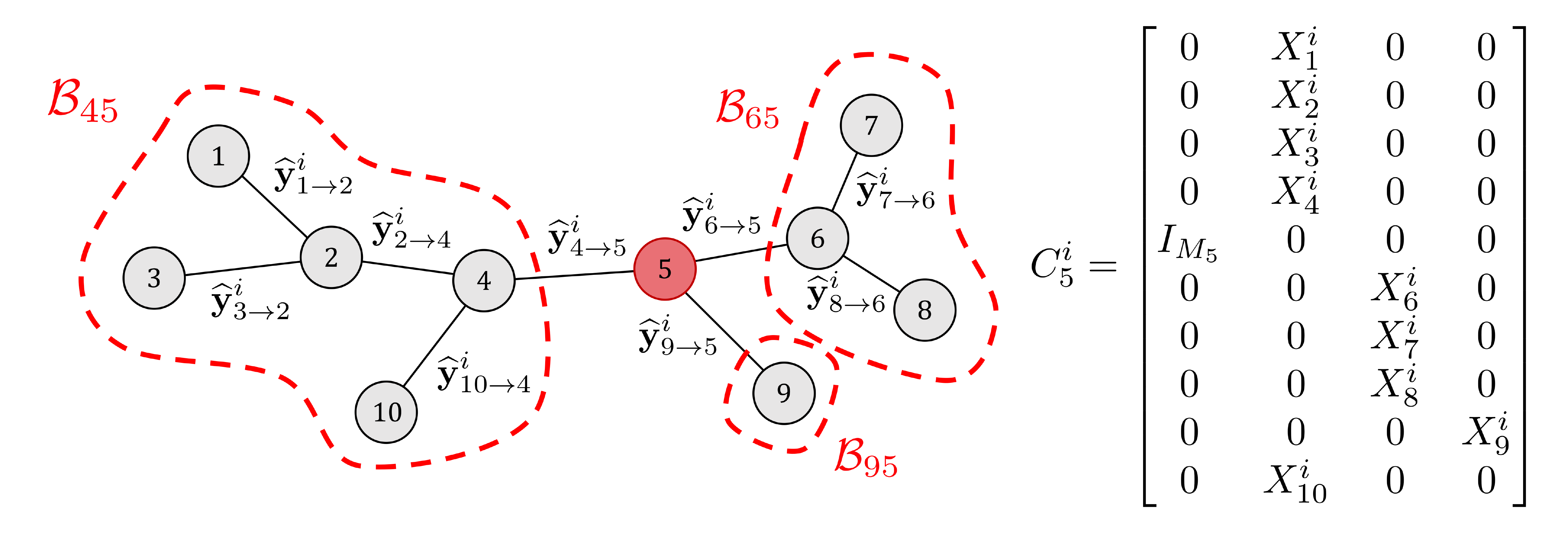}
  \caption{\cite{musluoglu2021distributed} Example of a tree network where the updating node is node $5$. Each neighbor of node $5$ creates its own cluster containing the nodes ``hidden'' from node $5$ behind them, shown here as $\mathcal{B}_{45}$, $\mathcal{B}_{65}$, $\mathcal{B}_{95}$. The resulting transition matrix is given by $C_5^i$.}
  \label{fig:tree_diagram}
\end{figure}

At the beginning of each iteration, every node compresses $N$ samples of its local signal $\mathbf{y}_k$ using its current estimate $X_k^i$ of the local variable $X_k$ to obtain the compressed $Q-$channel local signal:
\begin{equation}\label{eq:y_hat}
  \widehat{\mathbf{y}}^i_k(t)\triangleq X_k^{iT}\mathbf{y}_k(t)\in\mathbb{R}^{Q},
\end{equation}
while a similar operation is done to compress each $B_k$:
\begin{align}\label{eq:B_hat}
  \widehat{B}^i_k&\triangleq X_k^{iT}B_k\in\mathbb{R}^{Q\times L}.
\end{align}
A decentralized fuse-and-forward process is then started, in which the compressed data from all the nodes is summed on its way towards node $q$. This fuse-and-forward flow arises naturally when using a recursive computation rule, as explained in \cite{musluoglu2022unifiedp1}. Indeed, the data that node $k$ sends to its neighbor $n$ (which is closest to the updating node $q$) is defined as
\begin{equation}\label{eq:sum_fwd}
  \widehat{\mathbf{y}}_{k\rightarrow n}^i(t)\triangleq X_k^{iT}\mathbf{y}_k(t)+\sum_{l\in\mathcal{N}_k\backslash\{n\}}\widehat{\mathbf{y}}_{l\rightarrow k}^i(t),
\end{equation}
which can be computed as soon as node $k$ receives $\widehat{\mathbf{y}}_{l\rightarrow k}$ from all its neighbors $l\in\mathcal{N}_k$, except node $n$. We see that this recursive expression starts at leaf nodes (nodes with only one neighbor) and extends naturally to node $q$, which eventually receives
\begin{equation}\label{eq:sum_fwd_n}
  \widehat{\mathbf{y}}_{n\rightarrow q}^i(t)=X_n^{iT}\mathbf{y}_n(t)+\sum_{k\in\mathcal{N}_n\backslash\{q\}}\widehat{\mathbf{y}}_{k\rightarrow n}^i(t)=\sum_{k\in\mathcal{B}_{nq}}\widehat{\mathbf{y}}_k^i(t)
\end{equation}
from all its neighbors $n\in\mathcal{N}_q$. $\mathcal{B}_{nq}$ is the subgraph rooted at node $q$ that contains $n\in\mathcal{N}_q$, i.e., the set of nodes which would be disconnected from the subgraph containing node $q$ if we were to cut the connection between node $n$ and $q$ (see Figure \ref{fig:tree_diagram}). A similar fuse-and-forward process is performed for the terms (\ref{eq:B_hat}), resulting in fused parameters $\widehat{B}_{n\rightarrow q}^i$ \cite{musluoglu2022unifiedp1}.

The data received by node $q$ can then be structured in order to act as a local version of the global data. For this purpose, we define the $\widetilde{M}_q-$channel signal $\widetilde{\mathbf{y}}_q^i$ at node $q$ as the local signal $\mathbf{y}_q$ stacked with the compressed signals it receives from its neighbors, given by
\begin{equation}\label{eq:local_y}
  \widetilde{\mathbf{y}}_q^i(t)=[\mathbf{y}_q^T(t),\widehat{\mathbf{y}}_{n_1\rightarrow q}^{iT}(t),\dots,\widehat{\mathbf{y}}_{n_{|\mathcal{N}_q|}\rightarrow q}^{iT}(t)]^T.
\end{equation}
The matrix $\widetilde{B}_q^i$ is defined similarly.
Then, we define the local variable $\widetilde{X}_q$ at node $q$ as
\begin{equation}\label{eq:X_tilde_part}
  \widetilde{X}_q=[X_q^T,G_{n_1}^T,\dots,G_{n_{|\mathcal{N}_q|}}^T]^T\in\mathbb{R}^{\widetilde{M}_q\times Q},
\end{equation}
where $X_q\in\mathbb{R}^{M_q\times Q}$ and $G_n\in\mathbb{R}^{Q\times Q}$, $\forall n\in \mathcal{N}_q$.
This variable acts as a spatial fusion filter on the locally available data at node $q$, analogous to the way $X$ acts on $\mathbf{y}$ and $B$ for the global problem. From (\ref{eq:local_y}) and (\ref{eq:X_tilde_part}), we can then write
\begin{equation}
  \widetilde{X}_q^T\widetilde{\mathbf{y}}_q^i(t)=X_q^T\mathbf{y}_q(t)+\sum_{n\in\mathcal{N}_q}G_n^T\widehat{\mathbf{y}}_{n\rightarrow q}^i(t),
\end{equation}
and replacing the signals $\widehat{\mathbf{y}}_{n \rightarrow q}^i$ by their definition in (\ref{eq:sum_fwd_n}), we have
\begin{align}
  \widetilde{X}_q^T\widetilde{\mathbf{y}}_q^i(t)&=X_q^T\mathbf{y}_q(t)+\sum_{n\in\mathcal{N}_q}\sum_{k\in\mathcal{B}_{nq}}G_n^T\widehat{\mathbf{y}}_{k}^i(t),\\
  &=X_q^T\mathbf{y}_q(t)+\sum_{n\in\mathcal{N}_q}\sum_{k\in\mathcal{B}_{nq}}(X_k^iG_n)^T\mathbf{y}_{k}(t)\label{eq:Xytilde_sum}.
\end{align}
A similar expression can be derived for $\widetilde{X}_q^T\widetilde{B}_q^i$.

These local counterparts of the global expressions lead us to define the local problem at node $q$ using the previous parameterization of $X$, which is a compressed version of the global problem (\ref{eq:prob_g}):
\begin{equation}\label{eq:loc_prob_g}
  \begin{aligned}
    \underset{\widetilde{X}_q\in\mathbb{R}^{\widetilde{M}_q\times Q}}{\text{minimize } } \quad & \varphi(\widetilde{X}_q^T\widetilde{\mathbf{y}}_q^i(t),\widetilde{X}_q^T\widetilde{B}_q^i)\\
  \textrm{subject to} \quad & \eta_j(\widetilde{X}_q^T\widetilde{\mathbf{y}}_q^i(t),\widetilde{X}_q^T\widetilde{B}_q^i)\leq 0\;\textrm{ $\forall j\in\mathcal{J}_I$},\\
  & \eta_j(\widetilde{X}_q^T\widetilde{\mathbf{y}}_q^i(t),\widetilde{X}_q^T\widetilde{B}_q^i)=0\;\textrm{ $\forall j\in\mathcal{J}_E$}.
  \end{aligned}
\end{equation}
The fact that (\ref{eq:prob_g}) and (\ref{eq:loc_prob_g}) have an equivalent structure implies that the same solver can be used for both.

From (\ref{eq:Xytilde_sum}), we observe that the local inner product $\widetilde{X}_q^T\widetilde{\mathbf{y}}_q^i(t)$ at node $q$ is related to the network-wide inner product $X^T\mathbf{y}(t)$ if $X$ is defined as (details in \cite{musluoglu2022unifiedp1})
\begin{equation}\label{eq:X_to_X_tilde}
  X=C_q^i\widetilde{X}_q,
\end{equation}
with
\begin{equation}\label{eq:cqi_tree}
\begin{aligned}
&C_q(X)=\left[
    \begin{array}{c|c}
    0 &  \\
    I_{M_q} & \Theta_{-q}(X) \\
    0 & 
    \end{array}
    \right]\in\mathbb{R}^{M\times \widetilde{M}_q},\\
    &C_q^i\triangleq C_q(X^i),
\end{aligned}
\end{equation}
where $I_{M_q}$ is placed in the $q-$th block-row, and $\Theta_{-q}(X)$ is a block matrix with $K$ block-rows and $|\mathcal{N}_q|$ block-columns. An example of such a matrix $C_q^i$ is provided in Figure \ref{fig:tree_diagram}, which can be formally defined as follows. Each block-column corresponds to one of the neighbors $n\in\mathcal{N}_q$ of $q$, which we re-index as $m_n\in\{1,\dots,|\mathcal{N}_q|\}$. The block at the $k-$th block-row and $m_n-$th block-column is then defined as
\begin{equation}\label{eq:Theta_qi_def}
  \Big[\Theta_{-q}(X)\Big](k,m_n)=\begin{cases}
    X_k & \text{if $k\in\mathcal{B}_{nq}$} \\
    0 & \text{otherwise}.
    \end{cases}
\end{equation}
This transition matrix allows us to relate the global data or global variables with their local counterparts:
\begin{equation}\label{eq:y_B_X_tilde}
    \widetilde{\mathbf{y}}_q^i(t)=C_q^{iT}\mathbf{y}(t),\;\widetilde{B}_q^i=C_q^{iT}B,\;X=C_q^i \widetilde{X}_q
\end{equation}
and also write the local problem (\ref{eq:loc_prob_g}) in a compact way:
\begin{equation}\label{eq:loc_prob_g_compact}
  \begin{aligned}
  \underset{\widetilde{X}_q\in\mathbb{R}^{\widetilde{M}_q\times Q}}{\text{minimize } } \quad & f(C_q^i\widetilde{X}_q)\\
  \textrm{subject to} \quad & h_j(C_q^i\widetilde{X}_q)\leq 0\;\textrm{ $\forall j\in\mathcal{J}_I$,}\\
    & h_j(C_q^i\widetilde{X}_q)=0\;\textrm{ $\forall j\in\mathcal{J}_E$.}
  \end{aligned}
\end{equation}
Moreover, denoting the constraint set of the global problem (\ref{eq:prob_g}) or (\ref{eq:prob_g_compact}) as $\mathcal{S}$ and the constraint set of the local problem (\ref{eq:loc_prob_g}) or (\ref{eq:loc_prob_g_compact}) as $\widetilde{\mathcal{S}}_q^i$, it can be shown that (see \cite[Lemma 1]{musluoglu2022unifiedp1})
\begin{equation}\label{eq:loc_glob}
  \widetilde{X}_q\in\widetilde{\mathcal{S}}_q^i\iff C_q^i\widetilde{X}_q\in\mathcal{S},
\end{equation}
i.e., a point $\widetilde{X}_q$ in the constraint set of the local problem (\ref{eq:loc_prob_g}) leads to a corresponding point which by definition is also in the constraint set of the global problem (\ref{eq:prob_g}). Using this notation, we define the solution of the local problem (\ref{eq:loc_prob_g}) or equivalently (\ref{eq:loc_prob_g_compact}) as
\begin{align}\label{eq:comp_X_tilde}
  \widetilde{X}_q^{i+1}\triangleq\;&\underset{\widetilde{X}_q\in\widetilde{\mathcal{S}}_q^i}{\text{argmin }}f\left(C_q^i\widetilde{X}_q\right),\nonumber \\
  =\;&\underset{\widetilde{X}_q\in\widetilde{\mathcal{S}}_q^i}{\text{argmin }}\varphi\left(\widetilde{X}_q^T\widetilde{\mathbf{y}}_q^i(t),\widetilde{X}_q^T\widetilde{B}_q^i\right)
\end{align}
Considering instances where (\ref{eq:loc_prob_g}) or (\ref{eq:loc_prob_g_compact}) would have multiple global minima, we choose $\widetilde{X}_q^{i+1}$ as the solution minimizing the distance $||\widetilde{X}_q^{i+1}-\widetilde{X}_q^i||_F$, where
\begin{equation}\label{eq:X_fixed}
    \widetilde{X}_q^i\triangleq [X_q^{iT},I_Q,\dots,I_Q]^T.
\end{equation}

Finally, the matrices $G_{n_1}^{i+1},\dots,G_{n_{|\mathcal{N}_q|}}^{i+1}$ obtained from the partitioning (\ref{eq:X_tilde_part}) of $\widetilde{X}_q^{i+1}$ need to be disseminated in the network, so that every node can update their local estimator $X_k$ as
\begin{equation}\label{eq:upd_X}
  X_k^{i+1}=\begin{cases}
  X_q^{i+1} & \text{if $k=q$} \\
  X_k^{i}G_n^{i+1} & \text{if $k\neq q$, $k\in\mathcal{B}_{nq}$, $n\in\mathcal{N}_q$},
  \end{cases}
\end{equation}
which follows from the parameterization (\ref{eq:X_to_X_tilde}) of $X$.

\noindent This process is then repeated at a different node at each iteration, as summarized in Algorithm \ref{alg:dasf}.

\begin{figure}[!t]
  \removelatexerror
  \begin{algorithm}[H]
  \DontPrintSemicolon
  \caption{Distributed Adaptive Signal Fusion (DASF) Algorithm\;
  Code available in \cite{musluoglu2022dasftoolbox}}\label{alg:dasf}
  \SetKwInOut{Output}{output}
  \Output{$X^*$}
  \BlankLine
  Initialize $X^0$, $i\gets0$.\;
  \Repeat
  {
  Choose the updating node as $q\gets (i\mod K)+1$.\;
  1) The network $\mathcal{G}$ is pruned into a tree $\mathcal{T}^i(\mathcal{G},q)$.\;

  2) Every node $k$ collects a new batch of $N$ samples of $\mathbf{y}_k$. All nodes compress these to $N$ samples of $\widehat{\mathbf{y}}^{i}_k$ as in (\ref{eq:y_hat}). $\widehat{B}_k^i$ is computed using (\ref{eq:B_hat}).\;
  3) The nodes sum-and-forward their compressed data towards node $q$ via the recursive rule (\ref{eq:sum_fwd}) (and a similar rule for the $\widehat{B}_k^i$'s). Node $q$ eventually receives $N$ samples of $\widehat{\mathbf{y}}^{i}_{n\rightarrow q}$ given in (\ref{eq:sum_fwd_n}) along with $\widehat{B}_{n\rightarrow q}^i$ similarly defined, from all its neighbors $n\in\mathcal{N}_q$.\; 

  \At{Node q}
  {
    4a) Compute $\widetilde{X}_q^{i+1}$ as the solution of (\ref{eq:loc_prob_g}) where $\widetilde{\mathbf{y}}_q^i$, $\widetilde{B}_q^i$ and $\widetilde{X}_q^i$ are defined in (\ref{eq:local_y}) and (\ref{eq:X_tilde_part}). If the solution of (\ref{eq:loc_prob_g}) is not unique, select the solution which minimizes $||\widetilde{X}_q^{i+1}-\widetilde{X}_q^i||_F$ with $\widetilde{X}_q^i$ defined as in (\ref{eq:X_fixed}).\;

    4b) Partition $\widetilde{X}^{i+1}_q$ as in (\ref{eq:X_tilde_part}).\;
    4c) Disseminate $G_n^{i+1}$ to all nodes in $\mathcal{B}_{nq}$, $\forall n\in\mathcal{N}_q$.\;
  }
  
  5) Every node updates $X_k^{i+1}$ according to (\ref{eq:upd_X}).\;
  
  $i\gets i+1$\;
  }
  \end{algorithm}
  \small{\textbf{Note:} The fused output signal $\widehat{\mathbf{y}}(t)$ for the current batch of $N$ samples can be computed at node $q$ as $\widetilde{X}_q^{(i+1)T}\widetilde{\mathbf{y}}_q^i(t)$.}
  \hrule
\end{figure}

\section{Convergence and Optimality}\label{sec:conv_opt}
In this section, we analyze the convergence of the sequence of points $(X^i)_i$ that are generated by the DASF algorithm. Note that $X^i$ is formed by stacking all the local $X_k^i$'s at all nodes $k\in\mathcal{K}$, or equivalently, by using (\ref{eq:X_to_X_tilde}) which shows that $X^i=C_q^i\widetilde{X}_q^i$. We show that under some mild technical conditions (which are generally satisfied in practice), the DASF algorithm converges to a stationary point of (\ref{eq:prob_g}), and with some additional conditions even to the globally optimal signal fusion rule $X^*$, i.e., $(X^i)_i\rightarrow X^*\in\mathcal{X}^*$. As a result, the fused output signal $X^{iT}\mathbf{y}(t)$, which can be computed as $\widetilde{X}_k^{iT}\widetilde{\mathbf{y}}_k^i(t)$ at any node $k$, will become equal to $(X^*)^{T}\mathbf{y}(t)$ when the sample time $t$ is large (note that the iteration index $i$ is linked to the time index $t$). This convergence is without any of the nodes knowing the full signal $\mathbf{y}$ or matrix $B$, and despite the compression and fusion performed at each node.

For mathematical tractability, we make abstraction of estimation errors appearing from approximating the signal statistics using a finite $N$, which means the proofs only hold in the asymptotic case where $N\rightarrow +\infty$. In other words, we assume that all signal statistics are perfectly estimated, which is only an approximation of the practical case where the signal statistics are (re-)estimated based on blocks of $N$ samples. In practice, a finite $N$ has to be used, in which case the algorithm will hover around the optimal solution due to these aforementioned estimation errors in each iteration. We note that this is also the case for the centralized equivalent of the algorithm, in case the latter has to estimate the signal statistics over finite windows, e.g., in a tracking context.

\subsection{Convergence of the Objective}\label{sec:obj_conv}
The following result states the convergence of the objective function under Algorithm \ref{alg:dasf}'s update rule.
\begin{lem}\label{lem:monotonic}
  Let  $(X^i)_i$ be any sequence of iterates satisfying Algorithm \ref{alg:dasf}'s update rule for an instance $\mathbb{P}$ of (\ref{eq:prob_g}) or (\ref{eq:prob_g_compact}). Then, all $(X^i)_{i>0}$ belong to the constraint set $\mathcal{S}$ of (\ref{eq:prob_g_compact}) and $\left(f(X^i)\right)_{i>0}$ is a monotonically decreasing convergent sequence.
\end{lem}
\begin{proof}
Because $\widetilde{X}_q^{i+1}$ satisfies the constraint set $\widetilde{\mathcal{S}}_q^{i}$ (see (\ref{eq:comp_X_tilde})) for any update $i$, we conclude from (\ref{eq:loc_glob}) that each $X^i$, satisfies the constraint set $\mathcal{S}$ of the network-wide problem (\ref{eq:prob_g_compact}) for any $i\geq 1$ (as also shown in \cite[Lemma 1]{musluoglu2022unifiedp1}). Furthermore, since $\widetilde{X}_q^{i+1}$ is the solution of (\ref{eq:comp_X_tilde}), we have $f(C_q^i\widetilde{X}^{i+1}_q)\leq f(C_q^i\widetilde{X}_q)$ for any $\widetilde{X}_q\in\widetilde{\mathcal{S}}_q^i$. In particular, this inequality is verified for $\widetilde{X}_q=\widetilde{X}_q^i$ as defined in (\ref{eq:X_fixed}) for which $C_q^i\widetilde{X}_q^i=X^{i}$. This is because $\widetilde{X}_q^i$ indeed also belongs to the constraint set $\widetilde{\mathcal{S}}_q^i$ in (\ref{eq:comp_X_tilde}), because of (\ref{eq:loc_glob}) and the fact that $X^i$ belongs to $\mathcal{S}$ if $i \geq 1$ (see the beginning of the proof). Then, we have $f(C_q^i\widetilde{X}^{i+1}_q)=f(X^{i+1})\leq f(C_q^i\widetilde{X}_q^i)=f(X^i)$. Since the sequence is monotonically decreasing and since it has a lower bound (defined by the global minimum of $\mathbb{P}$), it must converge.
\end{proof}
\noindent Lemma \ref{lem:monotonic} guarantees that the sequence of objective values $\left(f(X^i)\right)_i$ associated with the iterates generated by the DASF algorithm converges and that the iterates correspond to feasible points of Problem (\ref{eq:prob_g}) or (\ref{eq:prob_g_compact}). However, this does not imply convergence of the sequence $(X^i)_i$ itself, which is typically much more challenging to guarantee, even for centralized optimization algorithms such as line-search or trust region methods \cite{absil2005convergence,goldstein1965steepest,fletcher2013practical,nocedal2006numerical,conn2000trust}. Moreover, even if the convergence of $(X^i)_i$ can be proven for the DASF algorithm, we still need to show that it converges to an ``interesting'' point, i.e., a stationary point of Problem (\ref{eq:prob_g_compact}), and preferably a global minimum. In the next two subsections, we will introduce two technical conditions (on top of the assumptions \ref{asmp:well_posed}-\ref{asmp:compactness} in Subsection \ref{sec:prob_descript}) which will be combined in Section \ref{sec:convergence_summary} to state convergence and optimality results for $(X^i)_i$. We end this subsection with a corollary of Lemma \ref{lem:monotonic} showing that the elements of $(X^i)_{i>0}$ lie in a compact set which is required to show convergence of the DASF algorithm (see Section \ref{sec:cond_conv} and Appendix \ref{app:continuity}).

\begin{cor}\label{cor:X0_compact}
 If either Assumption \ref{asmp:compact_sublevel} or \ref{asmp:compact_X0} is satisfied, then the elements of the sequence $(X^i)_{i>0}$ obtained from the DASF algorithm lie in a compact set.
\end{cor}
\noindent The proof of Corollary \ref{cor:X0_compact} comes directly from the monotonic decrease of the objective obtained in Lemma \ref{lem:monotonic}.

\subsection{Technical Conditions for Stationarity of Fixed Points}\label{sec:stat_cond}

The first condition comes in two versions, where either one of the two has to be satisfied in order to prove that the fixed points of the algorithm are stationary points. These conditions are akin to the linear independence constraint qualification (LICQ) in classical optimization theory \cite{peterson1973review}, and can be seen as compressed versions of these. We define a fixed point $\widebar{X}$ of the DASF algorithm as a point that is invariant under a DASF update step at any updating node $q$, i.e., $X^{i+1}=X^i=\widebar{X}$ independently of the updating node $q$ at iteration $i$.
\begin{subcond}{cond}\label{subcond:lin_indep}
\begin{cond}\label{cond:lin_indep_1}
  For a fixed point $\widebar{X}$ of Algorithm \ref{alg:dasf}, the elements of the set $\{\widebar{X}^{T}\nabla_X h_j(\widebar{X})\}_{j\in\mathcal{J}}$ are linearly independent.
\end{cond}
\noindent Since the set $\{\widebar{X}^{T}\nabla_X h_j(\widebar{X})\}_{j\in\mathcal{J}}$ consists of $J$ matrices of size $Q\times Q$, the number of constraints $J$ is upper bounded as
\begin{equation}\label{eq:J_upper_bound_1}
J\leq Q^2,
\end{equation}
in order to allow the set to be linearly independent.
As we will see in the proof of Theorem \ref{thm:stationarity}, this condition ensures that the Lagrange multipliers of the local problem are unique at a fixed point, eventually leading to the global optimality conditions being satisfied. 
Note that Condition \ref{cond:lin_indep_1} is highly likely to be satisfied in practice if (\ref{eq:J_upper_bound_1}) is satisfied, as a linear dependency would be highly coincidental if there are less matrices in the set than entries in each matrix (the points where this condition is violated is then a discrete set of points within a continuum of points). It is noted that Condition \ref{cond:lin_indep_1} can sometimes be shown to be satisfied a priori, based on the structure of the constraint set in the DSFO problem to which it is applied, without knowing the fixed points of the algorithm. The following example demonstrates how this can be proven when the constraint set is the Stiefel manifold, i.e., $X^TX=I_Q$, which is the case in, e.g., principal component analysis.

\begin{ex}\label{ex:stiefel}
Let $\mathcal{S}$ be the Stiefel manifold, i.e., we have the constraints $X^TX=I_Q$. There are $J=\frac{Q(Q-1)}{2}$ distinct constraints, where each constraint function is written as $h_{ml}(X)=\mathbf{x}(m)^T\mathbf{x}(l)-\mathbbm{1}\{m=l\}$, where $\mathbf{x}(m)$ is the $m-$th column of $X$, $m,l\in\{1,\dots,Q\}$, $l\leq m$ and $\mathbbm{1}$ is the indicator function. The derivative with respect to $X$ can be found to be
\begin{equation}
  \nabla_{X}h_{ml}(X)=\mathbf{x}(m)\mathbf{e}_l^T+\mathbf{x}(l)\mathbf{e}_m^T,
\end{equation}
where the $\mathbf{e}$'s are the standard basis vectors of $\mathbb{R}^Q$. Multiplying this expression by $X^T$ from the left and applying the constraint $X^TX=I_Q$ (assuming $X\in\mathcal{S}$), we have
\begin{equation}\label{eq:X_grad_h}
  X^T\nabla_{X}h_{ml}(X)=\mathbf{e}_m\mathbf{e}_l^T+\mathbf{e}_l\mathbf{e}_m^T.
\end{equation}
Note that the right-hand side of (\ref{eq:X_grad_h}) is independent of $X$. Since the $J$ elements of $\{\mathbf{e}_m\mathbf{e}_l^T+\mathbf{e}_l\mathbf{e}_m^T\}_{m,l}$, $l\leq m$, are linearly independent, Condition \ref{cond:lin_indep_1} is satisfied if $X\in\mathcal{S}$. From \cite[Lemma 1]{musluoglu2022unifiedp1}, every $X^i$ that is produced by Algorithm \ref{alg:dasf} belongs to $\mathcal{S}$, therefore Condition \ref{cond:lin_indep_1} is satisfied for $i>0$.
\end{ex}

Condition \ref{cond:lin_indep_1} allows us to state a first important result:
\begin{thm}\label{thm:stationarity}
Let $\mathbb{P}$ be an instance of (\ref{eq:prob_g}) or (\ref{eq:prob_g_compact}). Then, if Condition \ref{cond:lin_indep_1} is satisfied, any fixed point of Algorithm \ref{alg:dasf} must be a stationary point of $\mathbb{P}$ satisfying its KKT conditions.
\end{thm}
\begin{proof}
  See Appendix \ref{app:dasf_conv_1}.
\end{proof}
By definition, convergence of an algorithm can only be towards a fixed point of the algorithm, hence Theorem \ref{thm:stationarity} guarantees that \textit{if} the DASF algorithm converges, it converges to a stationary point of the global problem. For the special case of problems that are unconstrained, we do not require Condition \ref{cond:lin_indep_1} to hold (as it would lead to an empty set), in which case Theorem \ref{thm:stationarity} can be simplified as follows.
\begin{cor}\label{cor:stationarity_unconstrained}
Let $\mathbb{P}$ be an instance of (\ref{eq:prob_g}) or (\ref{eq:prob_g_compact}) which is unconstrained. Then, any fixed point of Algorithm \ref{alg:dasf} must be a stationary point of $\mathbb{P}$ satisfying its KKT conditions.
\end{cor}

Alternatively, we propose a less restrictive -- albeit more complicated -- condition for cases with more constraints than $Q^2$ (see (\ref{eq:J_upper_bound_1})), which is especially of interest for problems where $Q=1$, for which Condition \ref{cond:lin_indep_1} would only allow a single constraint.
\begin{cond}\label{cond:lin_indep_2}
For a fixed point $\widebar{X}$ of Algorithm \ref{alg:dasf}, the elements of the set $\{D_{j,q}(\widebar{X})\}_{j\in\mathcal{J}}$ are linearly independent for any $q$, where
\begin{equation}\label{eq:Dqj}
  D_{j,q}(\widebar{X})\triangleq \begin{bmatrix}
    \widebar{X}_q^{T}\nabla_{X_q} h_j(\widebar{X})\\
    \sum\limits_{k\in\mathcal{B}_{n_1q}}\widebar{X}_k^{T}\nabla_{X_k} h_j(\widebar{X})\\
    \vdots\\
    \sum\limits_{k\in\mathcal{B}_{n_{|\mathcal{N}_q|}q}}\widebar{X}_k^{T}\nabla_{X_k} h_j(\widebar{X})
  \end{bmatrix},
\end{equation}
which is a block-matrix containing $(1+|\mathcal{N}_q|)$ blocks of $Q\times Q$ matrices.
\end{cond}
\end{subcond}
\renewcommand*{\theHcond}{\thecond}
\noindent For a given node $q$, the elements of the set $\{D_{j,q}(\widebar{X})\}_{j\in\mathcal{J}}$ are now $(1+|\mathcal{N}_q|)Q\times Q$ matrices and therefore their size depends on the nodes and the topology of the network. This means that we require the number of constraints $J$ to satisfy:
\begin{equation}\label{eq:J_upper_bound_2_1}
  J\leq (1+\min_{k\in\mathcal{K}}|\mathcal{N}_k|)Q^2.
\end{equation}
This condition assumes that the pruning of the network $\mathcal{T}^i(\mathcal{G},q)$ preserves all the links with the neighbors of the updating node $q$. Furthermore, the proof in Appendix \ref{app:dasf_conv_2} will reveal that the number of constraints should also satisfy a second bound, which is also necessary for Condition \ref{cond:lin_indep_2} to hold:
\begin{equation}\label{eq:J_upper_bound_2_2}
    J\leq \frac{Q^2}{K-1}\sum_{k\in\mathcal{K}}|\mathcal{N}_k|.
\end{equation}
The reason is less obvious, but is related to specific interdependencies between the $D_{j,q}$'s across different nodes $q$ (see Appendix \ref{app:dasf_conv_2}). It is noted that both bounds (\ref{eq:J_upper_bound_2_1})-(\ref{eq:J_upper_bound_2_2}) are necessary, i.e., satisfying the first does not necessarily imply that the second one is satisfied and vice versa.

Similarly to the previous condition, Condition \ref{cond:lin_indep_2} is typically satisfied in practice when $J$ satisfies both bounds in (\ref{eq:J_upper_bound_2_1})-(\ref{eq:J_upper_bound_2_2}), i.e.,
\begin{equation}\label{eq:J_upper_bound_2}
    J\leq \min\left(\frac{Q^2}{K-1}\sum_{k\in\mathcal{K}}|\mathcal{N}_k|,\;(1+\min_{k\in\mathcal{K}}|\mathcal{N}_k|)Q^2\right).
\end{equation}
Nevertheless, it is still possible that there exists a fixed point that is ``close'' to violating this condition, in which case the convergence of the DASF algorithm might become very slow if it reaches a neighborhood of such a fixed point. We refer to Section \ref{sec:extra} on how to deal with these rare situations.

The following example illustrates why Condition \ref{cond:lin_indep_2} is less restrictive than Condition \ref{cond:lin_indep_1}.
\begin{ex}\label{ex:linear_constr}
Suppose that $\mathcal{S}=\{X\in\mathbb{R}^{M\times Q}\;|\;X^TB=A,\;A\in\mathbb{R}^{Q\times L}\}$, which is a typical constraint used in linearly constrained minimum variance (LCMV) beamforming \cite{van1988beamforming}. We have $J=QL$ constraints and each constraint function is given by $h_{ml}(X)=\mathbf{x}(m)^T\mathbf{b}(l)-A_{ml}$, where $\mathbf{x}(m)$ is the $m-$th column of $X$, $\mathbf{b}(l)$ is the $l-$th column of $B$ and $A_{ml}$ is the entry $(m,l)$ of the matrix $A$. It is straightforward to show that requiring the elements $X^T\nabla_X h_{ml}(X)=X^T\mathbf{b}(l)\mathbf{e}_m^T$ to be linearly independent for every $m$ and $l$, i.e., satisfying Condition \ref{cond:lin_indep_1}, is equivalent to requiring $\text{rank}(A)=L$. If $Q<L$, this condition cannot be satisfied. Looking now at Condition \ref{cond:lin_indep_2}, we have $X_k^T\nabla_{X_k} h_{ml}(X)=X_k^T\mathbf{b}_k(l)\mathbf{e}_m^T$ for every $k\in\mathcal{K}$, where $\mathbf{b}_k(l)$ is the block of $\mathbf{b}(l)$ corresponding to node $k$. Then, we can show that satisfying Condition \ref{cond:lin_indep_2} is equivalent to requiring that the matrix
\begin{equation}\label{eq:ex_lcmv_constr}
    [B_q^TX_q^{i},\sum_{k\in\mathcal{B}_{n_1q}}B_k^TX_k^{i},\dots,\sum_{k\in\mathcal{B}_{n_{|\mathcal{N}_q|}q}}B_k^TX_k^{i}]^T
\end{equation}
has rank $L$ at at a fixed point $X^i$, where the $B_k$'s are obtained from the partitioning of $B$ as in (\ref{eq:B_part}). This is possible even when $Q<L$, i.e., if node $q$ has sufficient neighbors such that $(1+|\mathcal{N}_q|)Q\geq L$.
\end{ex}

\begin{thm}\label{thm:stationarity_2}
Let $\mathbb{P}$ be an instance of (\ref{eq:prob_g}) or (\ref{eq:prob_g_compact}). Then, if Condition \ref{cond:lin_indep_2} is satisfied, a fixed point of Algorithm \ref{alg:dasf} must be a stationary point of $\mathbb{P}$ satisfying its KKT conditions.
\end{thm}
\begin{proof}
  See Appendix \ref{app:dasf_conv_2}.
\end{proof}

Finally, we note that these conditions are complementary in the sense that Condition \ref{cond:lin_indep_1} is not necessary for Condition \ref{cond:lin_indep_2} to hold, and vice versa.

\subsection{Technical Conditions for Convergence}\label{sec:cond_conv}
Conditions \ref{cond:lin_indep_1} and \ref{cond:lin_indep_2} are sufficient to show that fixed points of the DASF algorithm are stationary points. The next step is to show that accumulation points\footnote{We define an accumulation point of a sequence $(X^i)_{i\in\mathbb{N}}$ as the limit of a converging subsequence $(X^i)_{i\in\mathcal{I}}$ of $(X^i)_{i\in\mathbb{N}}$, with $\mathcal{I}\subseteq \mathbb{N}$.} of the algorithm are fixed points (and therefore stationary points of (\ref{eq:prob_g}) or (\ref{eq:prob_g_compact})), for which we require a second condition.

\begin{cond}\label{cond:continuity}
The local problems (\ref{eq:loc_prob_g}) or (\ref{eq:loc_prob_g_compact}) satisfy Assumptions \ref{asmp:well_posed}-\ref{asmp:compactness}.
\end{cond}

\noindent It is important to note here that this condition is usually satisfied in practice because the local problems have the same structure as the global problem (\ref{eq:prob_g}), which was already assumed to satisfy Assumptions \ref{asmp:well_posed}-\ref{asmp:compactness}. It is therefore reasonable to assume that these local problems also inherit these same properties. In Section \ref{sec:examples}, we will give several examples to illustrate how Condition \ref{cond:continuity} can be checked in various problems. We will also present some examples of rare cases where this condition is not satisfied and provide fixes for it.

The well-posedness of the problem as required in Assumption \ref{asmp:well_posed} requires a continuity assumption on the point-to-set mapping from the space of inputs of the problem to its solution space. Formally, let $\widetilde{\mathcal{F}}_q: \mathbb{R}^{M\times Q}\rightrightarrows \mathbb{R}^{\widetilde{M}_q\times Q}$ be the following point-to-set mapping:
\begin{equation}\label{def_FQ}
    \widetilde{\mathcal{F}}_q(X)\triangleq \underset{\widetilde{W}_q:C_q(X)\widetilde{W}_q\in\mathcal{S}_{q}(X)}{\text{argmin}} f(C_q(X)\widetilde{W}_q),
\end{equation}
where $C_q(X)$ is defined in (\ref{eq:cqi_tree}) and $\mathcal{S}_q$ is the point-to-set mapping corresponding to the local parameterized constraint set for $X$ when node $q$ is the updating node:
\begin{equation}\label{eq:pt_to_set_map}
    \mc{S}_q(X)=\{W\in\mc{S}\;|\;W_k\in\colspace{X_k}\quad \forall k\neq q\}
\end{equation}
with the subscript $k$ referring to the per-node partitioning of $X$ and $W$ as in (\ref{eq:X_part}) and $\colspace{X_k}$ the set of all matrices with the same size and the same column space as $X_k$. To appreciate how the set (\ref{eq:pt_to_set_map}) relates to the local constraint set, note that $\mathcal{S}_q(X^i)=\{X=C_q^i\widetilde{X}_q^i\;|\;\widetilde{X}_q^i\in\widetilde{\mathcal{S}}_q^i\}$. We require from our well-posedness property in Assumption \ref{asmp:well_posed} and Condition \ref{cond:continuity} that $\widetilde{\mathcal{F}}_q$ should be a continuous mapping (see \cite[Definition 17.2]{charalambos2013infinite} for a formal definition). Intuitively, this means that we expect that an arbitrarily small change in the inputs results in the addition or removal of points arbitrarily close to other points in the output set. In Section \ref{sec:examples}, we will illustrate on a few selected examples how the continuity of such a mapping can be checked.

Let the point-to-set mapping $\widetilde{\mathcal{M}}_q:\mathbb{R}^{M\times Q} \rightrightarrows \mathbb{R}^{\widetilde{M}_q\times Q}$ be defined as
\begin{equation}\label{eq:m_minimum_norm}
    \widetilde{\mathcal{M}}_q(X) \triangleq\underset{\widetilde{W}_q\in\widetilde{\mathcal{F}}_q(X)}{\text{argmin}} ||W_q-X_q||^2_F+\sum_{k\neq q} ||W_k-I_Q||^2_F,
\end{equation}
where $\widetilde{W}_q=[W_q^T,W_1^T,\dots,W_{q-1}^T,W_{q+1}^T,\dots,W_K^T]^T$, i.e., $\widetilde{\mathcal{M}}_q$ selects the point in the set $\widetilde{\mathcal{F}}_q(X)$ that is closest to $[X_q^T,I_Q,\dots,I_Q]^T$. We then define the point-to-set mapping $\mathcal{M}_q:\mathbb{R}^{M\times Q} \rightrightarrows \mathbb{R}^{M\times Q}$ as 
\begin{equation}\label{eq:m_minimum_norm2}
    {\mathcal{M}}_q(X) \triangleq C_q(X)  \widetilde{\mathcal{M}}_q(X).
\end{equation}
A single iteration of Algorithm \ref{alg:dasf} can then be summarized as
\begin{equation}\label{eq:alg}
    X^{i+1}\in\mathcal{M}_{q}(X^i).
\end{equation}
In very contrived cases, it could happen that $\mathcal{M}_q(X^i)$ is not a singleton, i.e., there exists more than one solution at a certain iteration $i$ which are equidistant to the previous estimate $X^i$. In that case, selecting by any means one particular solution is sufficient to resolve this ambiguity.

\begin{thm}\label{thm:cons_cont}
Suppose that for an instance $\mathbb{P}$ of (\ref{eq:prob_g}) or (\ref{eq:prob_g_compact}), under the updates of Algorithm \ref{alg:dasf}, Condition \ref{cond:continuity} is satisfied. Then:
\begin{enumerate}
    \item Any accumulation point $\widebar{X}$ of $(X^i)_i$ is a fixed point of the map $\mathcal{M}_q:\mathbb{R}^{M\times Q} \rightrightarrows \mathbb{R}^{M\times Q}$ for any $q$.
    \item $\lim_{i\rightarrow+\infty}||X^{i+1}-X^i||_F=0$.
\end{enumerate}
\end{thm}
\begin{proof}
  See Appendix \ref{app:continuity}.
\end{proof}

\noindent An important corollary is that any accumulation point is a fixed point of the full DASF algorithm as it is a fixed point for an update at any node $q$. However, note that Theorem \ref{thm:cons_cont} still does not guarantee convergence to a single point. The latter can be established if we assume the following condition:
\begin{subcond}{cond}\label{subcond:nb_stat_pts}
  \begin{cond}\label{cond:finite_stat}
    The number of stationary points of the global problem $\mathbb{P}$ is finite.
\end{cond}

\begin{thm}\label{thm:conv_single_pt}
If Conditions \ref{cond:continuity} and \ref{cond:finite_stat} are satisfied, then $(X^i)_i$ converges to a single point.
\end{thm}
\begin{proof}
  See Appendix \ref{app:continuity}.
\end{proof}

\noindent The condition on the finiteness of the number of stationary points can be relaxed to the following condition:

\begin{cond}\label{cond:finite_stat_solver}
The number of solutions of each local problem (\ref{eq:loc_prob_g_compact}) is finite or the solver of the local problems (\ref{eq:loc_prob_g_compact}) can only obtain a finite subset of the solutions of (\ref{eq:loc_prob_g_compact}).
\end{cond}
\end{subcond}
\renewcommand*{\theHcond}{\thecond}

In other words, we only require the finiteness of the number of solutions \textit{obtainable} through the solver used for solving the local problems. For example, when maximizing $\text{tr}(X^TAX)$ over $X^TBX=I$, there are infinitely many options for a solution $X^*$ represented as $X^*=V^*U$ where $U$ is an orthogonal matrix and $V^*$ contains the principal generalized eigenvectors of the matrix pencil $(A,B)$. However, a solver using a generalized eigenvalue decomposition to solve this problem can only output $V^*$ itself up to a sign change of its columns, hence the solver is only able to select solutions from a finite subset of the complete solution set.

\begin{thm}\label{thm:finite_stat_solver}
If Conditions \ref{cond:continuity} and \ref{cond:finite_stat_solver} are satisfied, then $(X^i)_i$ converges to a single point.
\end{thm}
\noindent The proof of Theorem \ref{thm:conv_single_pt} can be straightforwardly applied to Theorem \ref{thm:finite_stat_solver} as well.

\subsection{Convergence to Stationary Points and Global Minima}\label{sec:convergence_summary}
We can now combine all of the previous results to eventually obtain complete convergence results of the DASF algorithm to stationary points of Problem (\ref{eq:prob_g_compact}).

\begin{thm}\label{thm:dasf_conv_stat}
Suppose that for an instance $\mathbb{P}$ of (\ref{eq:prob_g}) or (\ref{eq:prob_g_compact}), under the updates of Algorithm \ref{alg:dasf}, Conditions \ref{subcond:lin_indep}, \ref{cond:continuity}, \ref{subcond:nb_stat_pts} (for \ref{subcond:lin_indep} and \ref{subcond:nb_stat_pts} either the form a or b) are satisfied, then $(X^i)_i$ converges and $\lim_{i\rightarrow+\infty}X^i=\widebar{X}$, where $\widebar{X}$ is a stationary point of $\mathbb{P}$ satisfying its KKT conditions.
\end{thm}

\begin{proof}
  From Theorems \ref{thm:conv_single_pt} and \ref{thm:finite_stat_solver}, $(X^i)_i$ converges to a single point $\widebar{X}$. Therefore $\widebar{X}$ is a fixed point of Algorithm \ref{alg:dasf} (Theorem \ref{thm:cons_cont}). From Theorems \ref{thm:stationarity} and \ref{thm:stationarity_2}, fixed points of the DASF algorithm are stationary points of $\mathbb{P}$ satisfying its KKT conditions, proving the theorem.
\end{proof}

Theorem \ref{thm:dasf_conv_stat} can lead to stronger convergence guarantees if all minima of the problem $\mathbb{P}$ are global minima (i.e., the value of the objective function is the same in all minima), which will be explained next. It is noted that many of the common spatial filtering design criteria satisfy this assumption, including PCA, canonical correlation analysis, minimum variance beamformers, generalized eigenvalue decomposition and trace ratio optimization.
\begin{thm}\label{thm:stability}
Under the same settings of Theorem \ref{thm:dasf_conv_stat}, if all minima of $\mathbb{P}$ are global minima, the only stable fixed points of Algorithm \ref{alg:dasf} are in $\mathcal{X}^*$.
\end{thm}
\begin{proof}
    For any fixed point $\widebar{X}\notin\mathcal{X}^*$, there exists a descent direction in $\mathcal{S}$ (as $\widebar{X}$ cannot be a minimum), and therefore there exists a perturbation $\Delta X$, with $X+\Delta X\in\mathcal{S}$, such that $f(X+\Delta X)\leq f(X)$. Due to the monotonic decrease of $(f(X^i))_i$ (see Lemma \ref{lem:monotonic}), the sequence $(X^i)_{i}$ is kicked out of equilibrium and cannot return to it, hence the equilibrium is unstable. Therefore, in the absence of local minima, the only stable fixed points of Algorithm \ref{alg:dasf} must be in $\mathcal{X}^*$.
\end{proof}

\begin{cor}\label{cor:global_minima}
  If all minima of $f$ over $\mathcal{S}$ are global minima and all conditions of Theorem \ref{thm:dasf_conv_stat} are satisfied, $(X^i)_i$ converges to the global minimum of (\ref{eq:prob_g}) or (\ref{eq:prob_g_compact}) with high probability\footnote{The phrasing ``with high probability'' here refers to the fact that it is expected that the algorithm cannot end up in an unstable equilibrium, as it would always escape from it due to numerical or estimation noise.}.
\end{cor}
\noindent This corollary follows immediately from Theorems \ref{thm:dasf_conv_stat} and \ref{thm:stability}. Indeed, from Theorem \ref{thm:dasf_conv_stat}, we know that $(X^i)_i$ converges to a stationary point. Since all fixed points that are not in $\mathcal{X}^*$ are unstable (Theorem \ref{thm:stability}), the algorithm will eventually escape from the neighborhoods of such unstable equilibria and will end up in a point of $\mathcal{X}^*$.

\begin{cor}\label{cor:strong_conv}
If the global problem (\ref{eq:prob_g}) or (\ref{eq:prob_g_compact}) is a convex problem with a strongly convex objective $f$ and all conditions of Theorem \ref{thm:dasf_conv_stat} are satisfied, $(X^i)_i$ converges to the unique global minimum $X^*$.
\end{cor}

\begin{proof}
  Theorem \ref{thm:dasf_conv_stat} guarantees convergence to a stationary point of Problem (\ref{eq:prob_g_compact}). The result comes from the fact that for a convex problem with a strongly convex objective, the unique stationary point is the global minimum.
\end{proof}

\noindent The previous corollary can be further relaxed by removing the requirements of Conditions \ref{cond:lin_indep_1} and \ref{cond:lin_indep_2} for a problem which is unconstrained (see Corollary \ref{cor:stationarity_unconstrained}):

\begin{cor}\label{cor:strong_conv_unconstrained}
If the global problem (\ref{eq:prob_g}) or (\ref{eq:prob_g_compact}) is unconstrained with a strongly convex objective $f$ and Condition \ref{cond:continuity} is satisfied, $(X^i)_i$ converges to the unique global minimum $X^*$.
\end{cor}

\section{Selected Examples}\label{sec:examples}

In this section, we illustrate how the different convergence results and conditions translate to some commonly encountered spatial filtering problems, and how the problems can be manipulated in order to satisfy the conditions in case they are violated. In the following, $\mathbf{y}$ and $\mathbf{d}$ are stochastic signals, and we denote their covariance and cross-covariance matrices $R_{\mathbf{yy}}=\mathbb{E}[\mathbf{y}(t)\mathbf{y}^T(t)]$ and $R_{\mathbf{yd}}=\mathbb{E}[\mathbf{y}(t)\mathbf{d}^T(t)]$. We also define the corresponding compressed matrices $R^i_{\widetilde{\mathbf{y}}_q\widetilde{\mathbf{y}}_q}=\mathbb{E}[\widetilde{\mathbf{y}}^i_q(t)\widetilde{\mathbf{y}}^{iT}_q(t)]=C_q^{iT}R_{\mathbf{yy}}C_q^i$ and $R^i_{\widetilde{\mathbf{y}}_q\mathbf{d}}=\mathbb{E}[\widetilde{\mathbf{y}}^i_q(t)\mathbf{d}^T(t)]=C_q^{iT}R_\mathbf{yd}$ for the local problems. 

As will be shown, we typically require $R_{\mathbf{yy}}$ to be non-singular for Condition \ref{cond:continuity} to be satisfied, in particular for the local problem to be well-posed (Assumption \ref{asmp:well_posed}), as Assumptions \ref{asmp:kkt} and \ref{asmp:compact_sublevel} (or \ref{asmp:compact_X0}) are satisfied automatically if the global problem satisfies these. We will see that a question that will arise is whether the local $R^i_{\widetilde{\mathbf{y}}_q\widetilde{\mathbf{y}}_q}$ is non-singular. If this is not satisfied, we cannot ignore situations where a small change in the local problem parameters leads to discontinuous changes in their solution sets, making Condition \ref{cond:continuity} (in particular Assumption \ref{asmp:well_posed}) invalid. The following result relates the (non-)singularity of $C_q^{iT}RC_q^i$ to the (non-)singularity of $R$.

\begin{lem}\label{lem:R_definite}
  Suppose that a matrix $R\in\mathbb{R}^{M\times M}$ is non-singular. Then, given $C_q^i\in\mathbb{R}^{M\times \widetilde{M}_q}$ as defined in (\ref{eq:cqi_tree}), if $C_q^i$ has full rank, the matrix $C_q^{iT}RC_q^i$ is also non-singular.
\end{lem}

\noindent The proof is omitted as this is a well-known property of the rank of a matrix. Lemma \ref{lem:R_definite} will be used in many of the examples that are discussed in this section.

\begin{rem}\label{rem:rank_Cq}
Note that --- by construction --- the matrix $C_q^i$ has full rank unless one of the matrices $X_k$ has linearly dependent columns. The latter is a contrived case and is to be avoided anyway, since it would imply that redundant data is transmitted via $\widehat{\mathbf{y}}_k$ (in which case the algorithm should only transmit the non-redundant part). The result of Lemma \ref{lem:R_definite} can therefore be interpreted in such a way that if the global problem is well-posed, then so is the local one because from this result, we have a guarantee that the local problems' covariance matrix will also be non-singular.
\end{rem}

\subsection{Least Squares / Minimum Mean Square Error and Ridge Regression}

The least squares (LS) / minimum mean square error (MMSE) problem can be written as
\begin{equation}\label{eq:prob_ls}
    \underset{X\in\mathbb{R}^{M\times Q}}{\text{minimize }}\mathbb{E}[||X^T\mathbf{y}(t)-\mathbf{d}(t)||^2],
\end{equation}
which is an unconstrained problem with a convex quadratic objective, since $R_{\mathbf{yy}}$ is positive semi-definite by definition. A solution $X^*$ of (\ref{eq:prob_ls}) needs to satisfy the normal equations given as $R_{\mathbf{yy}}X^*=R_{\mathbf{yd}}$. If additionally $R_{\mathbf{yy}}$ is positive definite, then it is invertible, and the objective is strongly convex leading to the unique solution $X^*=R_{\mathbf{yy}}^{-1}R_{\mathbf{yd}}$ of the global problem. In this case, (\ref{eq:prob_ls}) is well-posed and satisfies \textbf{Assumptions \ref{asmp:well_posed} and \ref{asmp:compactness}}, while \textbf{Assumption \ref{asmp:kkt}} is automatically satisfied as there are no constraints. We write the corresponding local problem (\ref{eq:loc_prob_g}) as
\begin{equation}\label{eq:loc_prob_ls}
    \underset{\widetilde{X}_q\in\mathbb{R}^{\widetilde{M}_q\times Q}}{\text{minimize }} \mathbb{E}[||\widetilde{X}_q^T\widetilde{\mathbf{y}}_q^i(t)-\mathbf{d}(t)||^2].
\end{equation}

\noindent Since (\ref{eq:loc_prob_ls}) does not have any constraints, \textbf{Condition \ref{cond:lin_indep_1}} is trivially satisfied (see also Corollary \ref{cor:stationarity_unconstrained}). Similarly to the centralized case, $\widetilde{X}_q$ is a solution of \eqref{eq:loc_prob_ls} if and only if it satisfies the normal equations
\begin{equation}
    R^i_{\widetilde{\mathbf{y}}_q\widetilde{\mathbf{y}}_q}\widetilde{X}_q=R^i_{\widetilde{\mathbf{y}}_q\mathbf{d}}.
\end{equation}

\noindent In general cases, $R^i_{\widetilde{\mathbf{y}}_q\widetilde{\mathbf{y}}_q}=C_q^{iT}R_{\mathbf{yy}}C_q^i$ is invertible from Lemma \ref{lem:R_definite}, and the solution of the local problem at iteration $i$ and node $q$ is unique (satisfying \textbf{Condition \ref{cond:finite_stat}}) and equal to
\begin{equation}
    \widetilde{X}^{i+1}_q=(R^i_{\widetilde{\mathbf{y}}_q\widetilde{\mathbf{y}}_q})^{-1}R^i_{\widetilde{\mathbf{y}}_q\mathbf{d}}.
\end{equation}
We then have $\widetilde{\mathcal{F}}_q(X^i)=\{\widetilde{X}_q^{i+1}\}$, which is continuous since matrix inversion is continuous (see Cramer's rule). Hence \textbf{Condition \ref{cond:continuity}} is satisfied if $R^i_{\widetilde{\mathbf{y}}_q\widetilde{\mathbf{y}}_q}$ is non-singular. Since (\ref{eq:prob_ls}) is unconstrained, we satisfy the conditions of \textbf{Corollary \ref{cor:strong_conv_unconstrained}}, and we conclude that the DASF algorithm applied to the LS / MMSE problem converges to the optimal solution $X^*$. 

In cases where $R^i_{\widetilde{\mathbf{y}}_q\widetilde{\mathbf{y}}_q}$ is singular, the normal equations have infinitely many solutions and (\ref{eq:loc_prob_ls}) therefore admits infinitely many solutions as well. However, a small change in inputs can lead to $R^i_{\widetilde{\mathbf{y}}_q\widetilde{\mathbf{y}}_q}$ suddenly becoming non-singular, reducing the number of solutions from infinitely many to a single one, which would not satisfy \textbf{Condition \ref{cond:continuity}}. We can resolve this problem by additionally requiring the column space of $\widetilde{X}_q$ to be orthogonal to the null space of $R^i_{\widetilde{\mathbf{y}}_q\widetilde{\mathbf{y}}_q}$, which corresponds to the solution with minimum norm \cite[Chapter 3, Section 2]{ben2003generalized}. The solution of this surrogate problem is always uniquely defined and varies continuously with $R^i_{\widetilde{\mathbf{y}}_q\widetilde{\mathbf{y}}_q}$ and $R^i_{\widetilde{\mathbf{y}}_q\mathbf{d}}$, even at points where $R^i_{\widetilde{\mathbf{y}}_q\widetilde{\mathbf{y}}_q}$ becomes singular. In practice though, $R^i_{\widetilde{\mathbf{y}}_q\widetilde{\mathbf{y}}_q}$ will never be singular (see also Remark \ref{rem:rank_Cq}), and the additional constraint is always trivially satisfied, making the surrogate problem equivalent to the original one.

If $R_{\mathbf{yy}}$ is singular (therefore implying that \textbf{Assumptions \ref{asmp:well_posed} and \ref{asmp:compactness}} do not hold in the general case for (\ref{eq:prob_ls})), one can consider adding an $\ell_2-$norm constraint or penalty to (\ref{eq:prob_ls}), leading to the ridge regression (RR) problem. As this can be rewritten as a least squares problem where $R_{\mathbf{yy}}$ is replaced by $R_{\mathbf{yy}}+\alpha I_M$, the resulting matrix becomes non-singular. In this case, both the local and global problems will satisfy the well-posedness assumption, and therefore \textbf{Corollary \ref{cor:strong_conv_unconstrained}} straightforwardly applies, such that convergence to the global solution is guaranteed.

\subsection{Linearly Constrained Minimum Variance}\label{sec:lcmv_example}

\begin{figure}[t]
  \includegraphics[width=0.48\textwidth]{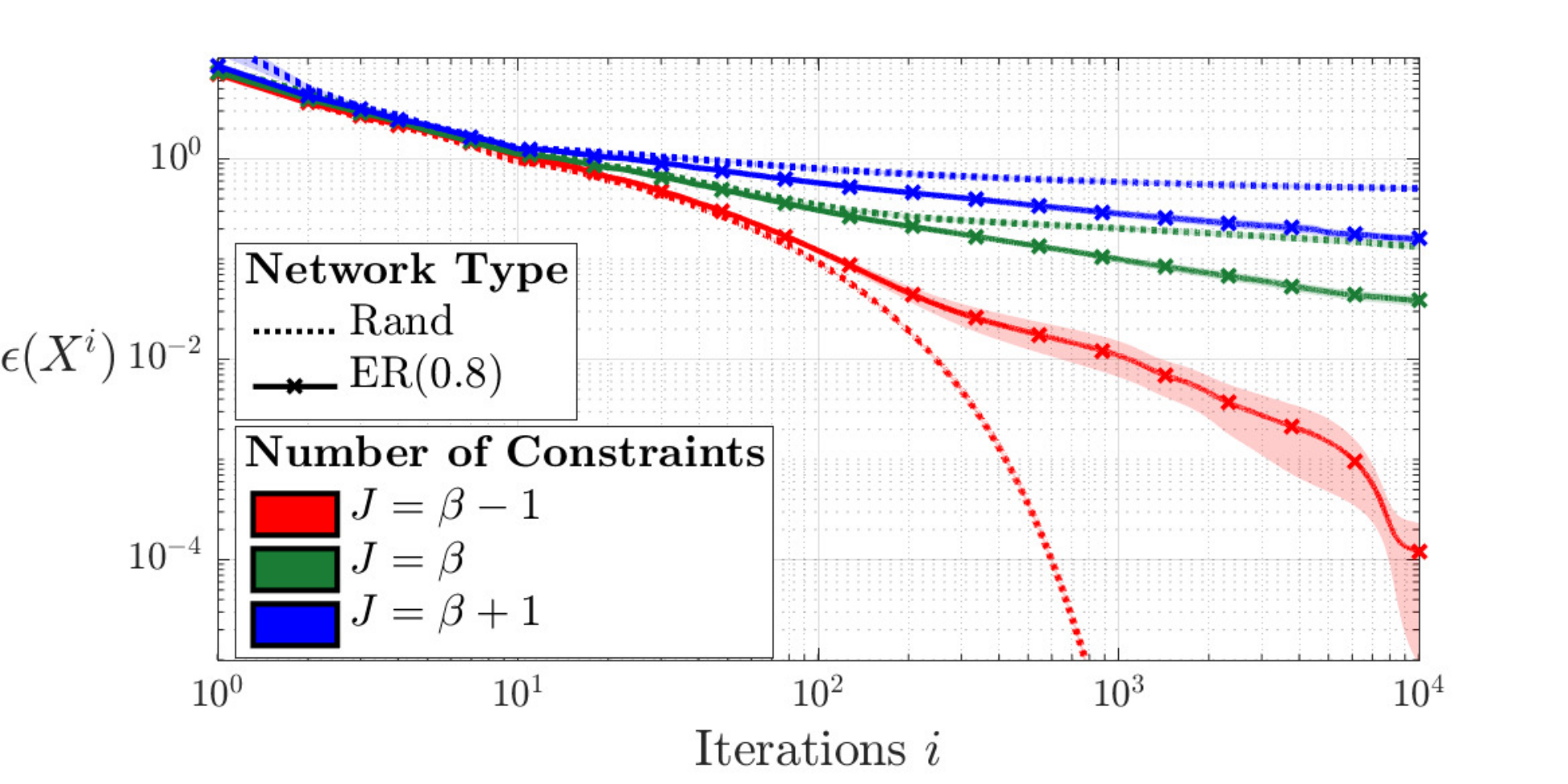}
  \caption{Convergence comparison of the DASF algorithm solving the LCMV problem (\ref{eq:prob_lcmv}) for Erd\H{o}s-R\'enyi random graphs with connection probability $0.8$ (\textit{ER($0.8$)}) and randomly generated trees (\textit{Rand}). The bold lines represent the mean values across $100$ Monte Carlo runs, while the shaded areas delimit the standard error of the mean around them. $\beta$ represents the upper bound given in (\ref{eq:J_upper_bound_2}) on the number of constraints $J$.}
  \label{fig:bound_plot}
\end{figure}

The linearly constrained minimum variance (LCMV) problem is a convex problem often used in beamforming applications \cite{van1988beamforming}, and written as
\begin{equation}\label{eq:prob_lcmv}
  \begin{aligned}
    \underset{X\in\mathbb{R}^{M\times Q}}{\text{minimize }} \quad &\mathbb{E}[||X^T\mathbf{y}(t)||^2]=\text{tr}(X^TR_{\mathbf{yy}}X)\\
  \textrm{subject to} \quad & X^TB=A,
  \end{aligned}
\end{equation}
with the linear term $B\in\mathbb{R}^{M\times L}$, $M>L$. If $R_{\mathbf{yy}}$ is positive definite and $\text{rank}(B)=L$, Problem (\ref{eq:prob_lcmv}) has a strongly convex objective and the unique solution is given by $X^*=R_{\mathbf{yy}}^{-1}B(B^TR_{\mathbf{yy}}^{-1}B)^{-1}A^T$. Problem (\ref{eq:prob_lcmv}) is then well-posed and satisfies \textbf{Assumption \ref{asmp:well_posed}}. As shown in Example \ref{ex:linear_constr}, the gradient of each constraint function $h_{ml}$ of (\ref{eq:prob_lcmv}) is given by $\nabla_X h_{ml}(X)=\mathbf{b}(l)\mathbf{e}_m$, where $\mathbf{b}(l)$ corresponds to the $l-$th column of $B$. Therefore, \textbf{Assumption \ref{asmp:kkt}} is satisfied when $B$ is full (column) rank. Finally, since the objective is continuous and the objective strongly convex, the sublevel sets of the objective are compact. Adding the constraints $X^TB=A$ preserves compactness as the intersection of a closed set with a compact one is compact, satisfying \textbf{Assumption \ref{asmp:compactness}}. The local LCMV problem is written as
\begin{equation}\label{eq:loc_prob_lcmv}
  \begin{aligned}
    \underset{\widetilde{X}_q\in\mathbb{R}^{\widetilde{M}_q\times Q}}{\text{minimize }} \quad &\text{tr}(\widetilde{X}_q^TR^i_{\widetilde{\mathbf{y}}_q\widetilde{\mathbf{y}}_q}\widetilde{X}_q)\\
  \textrm{subject to} \quad & \widetilde{X}_q^T\widetilde{B}_q^i=A,
  \end{aligned}
\end{equation}
at iteration $i$ and node $q$. As discussed in Example \ref{ex:linear_constr}, \textbf{Condition \ref{cond:lin_indep_1}} is satisfied if $A$ has rank $L$, or alternatively \textbf{Condition \ref{cond:lin_indep_2}} is satisfied if the matrix given in (\ref{eq:ex_lcmv_constr}) has rank $L$. Excluding the rare cases where $C_q^i$ is rank deficient (which can be dealt with, see Remark \ref{rem:rank_Cq}), $R^i_{\widetilde{\mathbf{y}}_q\widetilde{\mathbf{y}}_q}$ is invertible from Lemma \ref{lem:R_definite}, and it can be shown from the property that $B$ is full rank that $\widetilde{B}_q^{iT}(R^i_{\widetilde{\mathbf{y}}_q\widetilde{\mathbf{y}}_q})^{-1}\widetilde{B}_q^i$ is also invertible. Therefore, the unique solution of the local problem (implying \textbf{Condition \ref{cond:finite_stat}} is satisfied) is given by an analogous expression to the one of the global LCMV problem:
\begin{equation}
    \widetilde{X}_q^{i+1}=(R^i_{\widetilde{\mathbf{y}}_q\widetilde{\mathbf{y}}_q})^{-1}\widetilde{B}_q^i[\widetilde{B}_q^{iT}(R^i_{\widetilde{\mathbf{y}}_q\widetilde{\mathbf{y}}_q})^{-1}\widetilde{B}_q^i]^{-1}A^T.
\end{equation}
From the continuity of matrix inversion, \textbf{Condition \ref{cond:continuity}} is satisfied. This implies that we can apply \textbf{Corollary \ref{cor:strong_conv}} to conclude that the DASF algorithm will converge to the globally optimal LCMV solution. We note that \textbf{Condition \ref{cond:lin_indep_2}} is satisfied in practice with high probability when $J$ satisfies the upper bound (\ref{eq:J_upper_bound_2}) as illustrated in Figure \ref{fig:bound_plot}. However, there still exist situations where slow convergence is observed in cases where \textbf{Condition \ref{cond:lin_indep_2}} is ``close'' to being violated. We propose a fix for those situations in Section \ref{sec:split_nodes}. Note that Condition \ref{cond:lin_indep_2} is sufficient for showing convergence to the optimal point but it is not a necessary condition, as in Figure \ref{fig:bound_plot}, we can still see (slow) convergence for some cases when $J$ does not satisfy (\ref{eq:J_upper_bound_2}).

\subsection{Generalized Eigenvalue Decomposition and Principal Component Analysis}\label{sec:gevd_example}
Let us consider the problem:
\begin{equation}\label{eq:prob_gevd}
  \begin{aligned}
    \underset{X\in\mathbb{R}^{M\times Q}}{\text{minimize }} \quad -&\mathbb{E}\Big[||X^T\mathbf{y}(t)||^2\Big]=-\text{tr}(X^TR_{\mathbf{yy}}X)\\
  \textrm{subject to} \quad & \mathbb{E}[X^T\mathbf{v}(t)\mathbf{v}^T(t)X]=X^TR_{\mathbf{vv}}X=I_Q,
  \end{aligned}
\end{equation}
where $\mathbf{y}$ and $\mathbf{v}$ are $M-$dimensional time signals. Note that (\ref{eq:prob_gevd}) can be transformed into (\ref{eq:prob_g}) by minimizing the negative or the reciprocal of the objective. A global solution of this problem is obtained by computing the $Q$ principal generalized eigenvectors when computing the generalized eigenvalue decomposition (GEVD) of the matrix pencil $(R_{\mathbf{yy}},R_{\mathbf{vv}})$. This GEVD problem is often encountered in discriminant analysis or max-SNR filtering \cite{liu1992generalized,fukunaga2013introduction,van1988beamforming}. It also contains the standard eigenvalue decomposition (EVD) problem or principal component analysis (PCA) as a special case, which is obtained when the constraint set of (\ref{eq:prob_gevd}) is replaced with $X^TX=I_Q$, i.e., the Stiefel manifold (or equivalently, when $\mathbf{v}$ is a white noise process). Therefore, the discussions below apply to the EVD and PCA problems as well.

If both $R_{\mathbf{yy}}$ and $R_{\mathbf{vv}}$ are positive definite and the $Q+1$ largest generalized eigenvalues (GEVLs) of $(R_{\mathbf{yy}},R_{\mathbf{vv}})$ are all distinct, Problem (\ref{eq:prob_gevd}), is well-posed \cite{Kato2013}, therefore satisfying \textbf{Assumption \ref{asmp:well_posed}}. A solution of (\ref{eq:prob_gevd}) is given by $X^*=\text{GEVD}_Q(R_{\mathbf{yy}},R_{\mathbf{vv}})$, where we consider that, given a matrix pencil $(A,B)$, $\text{GEVD}_Q(A,B)$ is a matrix containing the $Q$ generalized eigenvectors of the pencil in its columns, corresponding to its largest GEVLs. We note that the solution of this problem is not unique, and applying any orthogonal transformation on $X^*=\text{GEVD}_Q(R_{\mathbf{yy}},R_{\mathbf{vv}})$ is also a valid solution. Similarly to Example \ref{ex:stiefel}, we have $\nabla_X h_{ml}(X^*)=R_{\mathbf{vv}}(\mathbf{x}^*(m)\mathbf{e}_l^T+\mathbf{x}^*(l)\mathbf{e}_m^T)$. It can be shown that the linear independence of the set $\{\nabla_X h_{ml}(X^*)\}_{m,l}$ is equivalent to the linear independence of the columns of $R_{\mathbf{vv}}X^*$. Under Assumption \ref{asmp:well_posed}, $R_{\mathbf{vv}}$ is positive definite, hence invertible and since $X^*$ contains generalized eigenvectors of $(R_{\mathbf{yy}},R_{\mathbf{vv}})$ in its columns, it is by definition full column rank. Therefore the solutions of (\ref{eq:prob_gevd}) satisfy the LICQ conditions hence \textbf{Assumption \ref{asmp:kkt}} is satisfied. Additionally, the sublevel sets of the objective of (\ref{eq:prob_gevd}) are closed, while the constraint of (\ref{eq:prob_gevd}) defines a compact set. From the compactness of their intersection, we satisfy \textbf{Assumption \ref{asmp:compactness}}. From (\ref{eq:prob_gevd}), we observe that the corresponding local problem (\ref{eq:loc_prob_g}) is given by
\begin{equation}\label{eq:loc_prob_gevd}
  \begin{aligned}
    \underset{\widetilde{X}_q\in\mathbb{R}^{\widetilde{M}_q\times Q}}{\text{minimize }} \quad -&\text{tr}(\widetilde{X}_q^TR^i_{\widetilde{\mathbf{y}}_q\widetilde{\mathbf{y}}_q}\widetilde{X}_q)\\
  \textrm{subject to} \quad & \widetilde{X}_q^TR^i_{\widetilde{\mathbf{v}}_q\widetilde{\mathbf{v}}_q}\widetilde{X}_q=I_Q,
  \end{aligned}
\end{equation}
and for any fixed iteration $i$ and node $q$, a solution of the local problem is
\begin{equation}\label{eq:gevd_sol}
    \widetilde{X}_q^{i+1}=\text{GEVD}_Q(R^i_{\widetilde{\mathbf{y}}_q\widetilde{\mathbf{y}}_q},\;R^i_{\widetilde{\mathbf{v}}_q\widetilde{\mathbf{v}}_q}).
\end{equation}
Similarly to Example \ref{ex:stiefel}, we can show that $\nabla_X h_{ml}(X)=R_{\mathbf{vv}}(\mathbf{x}(m)\mathbf{e}_l^T+\mathbf{x}(l)\mathbf{e}_m^T)$. Therefore, for any $X$ satisfying the constraints of (\ref{eq:prob_gevd}), we have $X^T\nabla_X h_{ml}(X)=\mathbf{e}_m\mathbf{e}_l^T+\mathbf{e}_l\mathbf{e}_m^T$, which, for every $(m,l)$, form a linearly independent set hence \textbf{Condition \ref{cond:lin_indep_1}} is satisfied at any iteration.

On the other hand, we do not have a guarantee that the algorithm does not converge to a local problem where the $Q+1$ largest generalized eigenvalues of the local matrix pencil are all distinct. This would lead to a violation of the continuity of the problem as required in \textbf{Condition \ref{cond:continuity}}, which is otherwise satisfied. This event, however improbable, can be monitored and a particular fix is described in Subsection \ref{sec:non_cont_loc}.

As noted previously, there exists infinitely many solutions of Problem (\ref{eq:prob_gevd}) therefore \textbf{Condition \ref{cond:finite_stat}} cannot be satisfied. However, suppose the solver we choose to solve the local problems (\ref{eq:loc_prob_gevd}) computes the generalized eigenvalue decomposition of the matrix pencil $(R^i_{\widetilde{\mathbf{y}}_q\widetilde{\mathbf{y}}_q},\;R^i_{\widetilde{\mathbf{v}}_q\widetilde{\mathbf{v}}_q})$ as in (\ref{eq:gevd_sol}). Then, the solver can only output one of the $2^Q$ possible solutions of the local problems at each iteration, namely one of the matrices containing the $Q$ most significant generalized eigenvectors of the matrix pencil $(R^i_{\widetilde{\mathbf{y}}_q\widetilde{\mathbf{y}}_q},\;R^i_{\widetilde{\mathbf{v}}_q\widetilde{\mathbf{v}}_q})$, which are equal up to a sign change of the columns. This allows to eliminate all other solutions of (\ref{eq:loc_prob_gevd}) from the set of candidate solutions, making the solution set obtainable by the solver finite and leading to \textbf{Condition \ref{cond:finite_stat_solver}} being satisfied. 

Under these conditions, we conclude from \textbf{Theorem \ref{thm:dasf_conv_stat}} that the DASF algorithm converges to a stationary point of the GEVD problem. As all minima/maxima of (\ref{eq:prob_gevd}) are global minima/maxima, we obtain convergence to the global solution according to \textbf{Corollary \ref{cor:global_minima}}.

\subsection{Trace Ratio Optimization}
The trace ratio optimization (TRO) problem \cite{wang2007trace,ngo2012trace} is defined as
\begin{equation}\label{eq:prob_tro}
  \begin{aligned}
    \underset{X\in\mathbb{R}^{M\times Q}}{\text{minimize }} \quad -&\frac{\mathbb{E}\Big[||X^T\mathbf{y}(t)||^2\Big]}{\mathbb{E}\Big[||X^T\mathbf{v}(t)||^2\Big]}=-\frac{\text{tr}(X^TR_{\mathbf{yy}}X)}{\text{tr}(X^TR_{\mathbf{vv}}X)}\\
  \textrm{subject to} \quad & X^TX=I_Q.
  \end{aligned}
\end{equation}
Considering that $R_{\mathbf{yy}}$ and $R_{\mathbf{vv}}$ are positive definite, there exists a scalar $\rho$ such that a solution of this problem is given by $X^*=\text{EVD}_Q(R_{\mathbf{yy}}-\rho R_{\mathbf{vv}})$ \cite{wang2007trace}, where $\text{EVD}_Q(A)$ is a matrix containing the $Q$ eigenvectors corresponding to the $Q$ largest eigenvalues of $A$ in its columns. This solution returned by the TRO solver defined in \cite{wang2007trace} is unique up to a sign change of its columns if the $Q+1$ eigenvalues of $R_{\mathbf{yy}}-\rho R_{\mathbf{vv}}$ are distinct, in which case the problem is well-posed, satisfying \textbf{Assumption \ref{asmp:well_posed}}. It can be shown in a similar fashion as to \ref{sec:gevd_example} that (\ref{eq:prob_tro}) satisfies \textbf{Assumptions \ref{asmp:kkt} and \ref{asmp:compactness}}. Additionally, From Example \ref{ex:stiefel}, we know that the Stiefel manifold satisfies \textbf{Condition \ref{cond:lin_indep_1}}.
The local problem that node $q$ solves at iteration $i$ is written as
\begin{equation}\label{eq:loc_prob_tro}
  \begin{aligned}
    \underset{\widetilde{X}_q\in\mathbb{R}^{\widetilde{M}_q\times Q}}{\text{minimize }} \quad -&\frac{\text{tr}(\widetilde{X}_q^TR^i_{\widetilde{\mathbf{y}}_q\widetilde{\mathbf{y}}_q}\widetilde{X}_q)}{\text{tr}(\widetilde{X}_q^TR^i_{\widetilde{\mathbf{v}}_q\widetilde{\mathbf{v}}_q}\widetilde{X}_q)}\\
  \textrm{subject to} \quad & \widetilde{X}_q^T\widetilde{\Gamma}_q^i\widetilde{X}_q=I_Q,
  \end{aligned}
\end{equation}
with\footnote{The expression $X^TX$ can be rewritten in the form required by $Z_B$ in (\ref{eq:prob_g}) if $B$ is set to $B=I_M$, which results in $(X^TB)\cdot (X^TB)^T$. It is then observed from (\ref{eq:y_B_X_tilde}) that $\widetilde{B}_q^i=C_q^i$.} $\widetilde{\Gamma}_q^i=C_q^{iT}C_q^i$. There exists a scalar $\rho_q^i$ such that the solution of the local problem is given by
\begin{equation}\label{eq:loc_prob_gevd_X}
    \widetilde{X}_q^{i+1}=\text{GEVD}_Q\left(R^i_{\widetilde{\mathbf{y}}_q\widetilde{\mathbf{y}}_q}-\rho^i_qR^i_{\widetilde{\mathbf{v}}_q\widetilde{\mathbf{v}}_q},\;\widetilde{\Gamma}_q^i\right),
\end{equation}
if both matrices of the pencil are positive definite (this holds if $C_q^i$ is full rank from Lemma \ref{lem:R_definite}) and the $Q+1$ largest GEVLs are distinct (in Subsection \ref{sec:non_cont_loc}, we will cover the case where this is not satisfied). Therefore, the local problems are generally well-posed, satisfying \textbf{Condition \ref{cond:continuity}}. Similarly to the GEVD example, a solver using (\ref{eq:loc_prob_gevd_X}) to solve (\ref{eq:loc_prob_tro}) satisfies \textbf{Condition \ref{cond:finite_stat_solver}}. As there are no local minima, in practical cases we obtain convergence to a global minimum from \textbf{Corollary \ref{cor:global_minima}}.
\subsection{Canonical Correlation Analysis}
The goal of canonical correlation analysis (CCA) is to find two spatial filters for two different multi-channel signals such that their outputs are maximally correlated \cite{borga1998learning,hardoon2004canonical}. The CCA problem can be (re)written as \cite{hardoon2004canonical}
\begin{equation}\label{eq:prob_cca}
  \begin{aligned}
    \underset{X,W\in\mathbb{R}^{M\times Q}}{\text{minimize }} \quad -&\mathbb{E}\Big[\text{tr}\left(X^T\mathbf{y}(t)\mathbf{v}^T(t)W\right)\Big]=-\text{tr}(X^TR_{\mathbf{yv}}W)\\
  \textrm{subject to} \quad & \mathbb{E}[X^T\mathbf{y}(t)\mathbf{y}^T(t)X]=X^TR_{\mathbf{yy}}X=I_Q,\\
  & \mathbb{E}[W^T\mathbf{v}(t)\mathbf{v}^T(t)W]=W^TR_{\mathbf{vv}}W=I_Q.
  \end{aligned}
\end{equation}
Assuming that $R_{\mathbf{yy}}$ and $R_{\mathbf{vv}}$ are positive definite the solution of Problem (\ref{eq:prob_cca}) is given by $X^*=\text{GEVD}_Q(R_{\mathbf{yv}}R_{\mathbf{vv}}^{-1}R_{\mathbf{vy}},\;R_{\mathbf{yy}})$ and $W^*=R_{\mathbf{vv}}^{-1}R_{\mathbf{vy}}X^*\Lambda^{-1/2}$, where $\Lambda$ is a $Q\times Q$ diagonal matrix containing the $Q$ largest GEVLs of the pencil $(R_{\mathbf{yv}}R_{\mathbf{vv}}^{-1}R_{\mathbf{vy}},\;R_{\mathbf{yy}})$. If the $Q+1$ largest GEVLs of this pencil are all distinct, the CCA problem is well-posed, satisfying \textbf{Assumption \ref{asmp:well_posed}}. Similarly to \ref{sec:gevd_example}, it can be shown that (\ref{eq:prob_cca}) satisfies \textbf{Assumptions \ref{asmp:kkt} and \ref{asmp:compactness}}.
As shown previously, \textbf{Condition \ref{cond:lin_indep_1}} is satisfied at any iteration $i$. We can write the local problems as
\begin{equation}\label{eq:loc_prob_cca}
  \begin{aligned}
    \underset{\widetilde{X}_q,\widetilde{W}_q\in\mathbb{R}^{\widetilde{M}_q\times Q}}{\text{minimize }} \quad -&\text{tr}(\widetilde{X}_q^TR^i_{\widetilde{\mathbf{y}}_q\widetilde{\mathbf{v}}_q}\widetilde{W}_q)\\
  \textrm{subject to} \quad & \widetilde{X}_q^TR^i_{\widetilde{\mathbf{y}}_q\widetilde{\mathbf{y}}_q}\widetilde{X}_q=I_Q,\\
  & \widetilde{W}_q^TR^i_{\widetilde{\mathbf{v}}_q\widetilde{\mathbf{v}}_q}\widetilde{W}_q=I_Q.
  \end{aligned}
\end{equation}
From Lemma \ref{lem:R_definite} and Remark \ref{rem:rank_Cq}, we conclude that both $R^i_{\widetilde{\mathbf{y}}_q\widetilde{\mathbf{y}}_q}$ and $R^i_{\widetilde{\mathbf{v}}_q\widetilde{\mathbf{v}}_q}$ are non-singular. The solution of the local problems is then
\begin{align}
    \widetilde{X}_q^{i+1}&=\text{GEVD}_Q\left(R^i_{\widetilde{\mathbf{y}}_q\widetilde{\mathbf{v}}_q}(R^i_{\widetilde{\mathbf{v}}_q\widetilde{\mathbf{v}}_q})^{-1}R^i_{\widetilde{\mathbf{v}}_q\widetilde{\mathbf{y}}_q},\;R^i_{\widetilde{\mathbf{y}}_q\widetilde{\mathbf{y}}_q}\right) \label{eq:prob_loc_cca_X}\\
    \widetilde{W}_q^{i+1}&=(R^i_{\widetilde{\mathbf{v}}_q\widetilde{\mathbf{v}}_q})^{-1}R^i_{\widetilde{\mathbf{v}}_q\widetilde{\mathbf{y}}_q}\widetilde{X}_q^{i+1}(\Lambda_q^i)^{-1/2}\label{eq:prob_loc_cca_W},
\end{align}
where $\Lambda_q^i$ is the $Q\times Q$ diagonal matrix containing the $Q$ largest GEVLs of the pencil given in (\ref{eq:prob_loc_cca_X}). If the $Q+1$ GEVLs are all distinct, the local problem (\ref{eq:loc_prob_cca}) is again well-posed and \textbf{Condition \ref{cond:continuity}} is satisfied. If this is not the case, the same fixes we discussed in the examples of the GEVD and TRO problems can be applied. As in these latter examples, \textbf{Condition \ref{cond:finite_stat_solver}} is also satisfied when choosing a solver which computes the solutions of (\ref{eq:loc_prob_cca}) using (\ref{eq:prob_loc_cca_X})-(\ref{eq:prob_loc_cca_W}). Therefore, the corresponding DASF algorithm will converge to the centralized CCA solution from \textbf{Corollary \ref{cor:global_minima}}.
\section{Fixes in Case of Violations of Conditions}\label{sec:extra}

\subsection{Avoiding Violations of Condition \ref{cond:lin_indep_2}}\label{sec:split_nodes}
In certain cases, the linear independence requirements of Condition \ref{cond:lin_indep_1} or \ref{cond:lin_indep_2} cannot be a priori guaranteed. In such cases, even though they are expected to hold (since they are only violated in a discrete set of points within a continuum of possibilities), it is possible that the DASF algorithm passes a neighborhood of a fixed point $\widebar{X}$ where the conditions are not satisfied, especially when $J$ is close to the upper bound (\ref{eq:J_upper_bound_2}) or even larger (see Figure \ref{fig:bound_plot}). 
When an updating node $q$ observes that the elements of $\{D_{j,q}(X^i)\}_j$ are close to being linearly dependent, the algorithm is potentially converging to a fixed point that does not satisfy the linear independence requirements. This can be checked at the node itself, e.g., when the $J-$th singular value of $[\text{vec}(D_{1,q}(X^i)),\dots,\text{vec}(D_{J,q}(X^i))]$ is close to zero\footnote{Note that, depending on the problem, constructing $D_{j,q}$'s at node $q$ might require some additional data exchange, namely each node $k$ should compute and transmit $X_k^{iT}\nabla_{X_k}h_j(X^i)$'s towards node $q$. However, this would not add a significant burden to the communication cost as these matrices are $Q\times Q$, while there are also many cases where the compressed gradients are already required to be communicated due to the problem's structure.}. In that case, we propose that the neighbors $k$ of node $q$ split their local $X_k^i$ into a sum of two matrices $X_{k,a}^i$ and $X_{k,b}^i$ (which both have linearly independent columns) such that $X_k^i=X_{k,a}^i+X_{k,b}^i$, and start to temporarily communicate two sets of compressed signals $\widehat{\mathbf{y}}_{k,a}^i=X_{k,a}^{iT}\mathbf{y}_k$ and $\widehat{\mathbf{y}}_{k,b}^i=X_{k,b}^{iT}\mathbf{y}_k$.
This implies that the elements of $\{D_{j,q}(X^i)\}_j$ have grown in dimension thereby making the linear independence requirement of the set $\{D_{j,q}(X^i)\}_j$ more likely to be met. As soon as the algorithm escapes the suboptimal point, node $k$ can again merge $X_{k,a}$ and $X_{k,b}$ for future iterations, returning to the minimal communication bandwidth setting.

\subsection{Avoiding Violations of Condition \ref{cond:continuity}}\label{sec:non_cont_loc}
In the very contrived cases where the algorithm would produce a subsequence converging to a stationary point at which Condition \ref{cond:continuity} does not hold, convergence of the overall sequence cannot be guaranteed anymore and the algorithm will possibly oscillate between points which are solutions of the local problems, but not necessarily corresponding to global solutions. Indeed, without Condition \ref{cond:continuity}, we cannot guarantee that $\lim_{X\to \widebar{X}}\mathcal{M}_q(X)$ is well-defined (i.e., is unique and a singleton). A practical and pragmatic fix for such scenarios is to monitor both the potential oscillatory behavior and the continuity of $\widetilde{\mathcal{F}}_q(X^i)$ and skip the update at the node where a problem occurs. To detect such an oscillation, select an arbitrarily small $\varepsilon>0$, and monitor whether $|f(X^{i+1})-f(X^i)|\cdot||X^{i+1}-X^i||_F^{-1}>\varepsilon$ which can be interpreted as a sufficient decrease condition. If this condition is violated, a potential oscillation is flagged. In addition, one could monitor whether a particular near-discontinuity condition is met, which is problem specific. This condition can be interpreted as a sufficient decrease condition. In the case of GEVD and TRO problems discussed above, monitoring the difference between the $Q-$th and $(Q+1)-$th largest eigenvalue is sufficient to detect such a discontinuity. If both such a near-discontinuity and an insufficient decrease are flagged, the update at that node should be skipped.
\section{Conclusion}\label{sec:conclusion}
In this paper, we have analyzed the technical convergence properties and conditions of the DASF algorithm and provided formal proofs of convergence. The conditions required for convergence were shown to be satisfied in many practical problems, assuming some bounds on the dimension of the problem are satisfied, which depend on the number of constraints and the network topology. We have provided some illustrative examples to demonstrate how the --- sometimes rather technical --- conditions can be validated in practice. These examples also showed in which contrived cases a problem could occur, e.g., in case of singularities or eigenvalue collisions in sensor signal covariance matrices, which are expected to be rare in practice. Nevertheless, we have discussed various methods to fix these convergence problems for the rare cases where these should occur.
\appendices
\section{ }\label{app:dasf_conv_1}

\begin{proof}[Proof of Theorem \ref{thm:stationarity}]
Let us write the KKT conditions of Problem (\ref{eq:prob_g_compact}), as mentioned in Assumption \ref{asmp:kkt} \cite{bertsekas1999nonlinear}:
\begin{align}
  &\nabla_X \mathcal{L}(X,\bm{\lambda})=0,\label{eq:stat}\\
  &h_j(X)\leq 0\;\textrm{ $\forall j\in\mathcal{J}_I$, } h_j(X)=0\;\textrm{ $\forall j\in\mathcal{J}_E$,}\label{eq:pf}\\
  &\lambda_j\geq 0\;\textrm{ $\forall j\in\mathcal{J}_I$,}\label{eq:dual}\\
  &\lambda_j h_j(X)=0\;\textrm{ $\forall j\in\mathcal{J}_I$,}\label{eq:comp_slack}
\end{align}
where 
\begin{equation}
  \mathcal{L}(X,\bm{\lambda})\triangleq f(X)+\sum_{j\in\mathcal{J}} \lambda_jh_j(X)
\end{equation}
is the Lagrangian with $\lambda_j\in\mathbb{R}$ the Lagrange multiplier corresponding to the constraint $h_j$ and $\bm{\lambda}$ in bold is used as a shorthand notation for the set of all Lagrange multipliers. These KKT conditions can also be formalized for the local optimization problem (\ref{eq:loc_prob_g_compact}) defined at the updating node $q$. Since $\widetilde{X}_q^{i+1}$ solves the local problem (\ref{eq:loc_prob_g_compact}), it must satisfy the KKT conditions of the local problem, and therefore based on the parameterization in (\ref{eq:loc_prob_g_compact}), the stationarity condition can be written as
\begin{equation}\label{eq:optimality_local}
  \nabla_{\widetilde{X}_q}\mathcal{L}(C_q^i\widetilde{X}_q^{i+1},\widetilde{\bm{\lambda}}(q))=0,
\end{equation}
where $\widetilde{\bm{\lambda}}(q)$ represents the set of Lagrange multipliers corresponding to the local problem (\ref{eq:loc_prob_g_compact}) at node $q$ and iteration $i$. Note that (\ref{eq:optimality_local}) contains the Lagrangian of the centralized problem (\ref{eq:prob_g_compact}), yet it is parameterized based on (\ref{eq:loc_prob_g_compact}) to transform it into the local problem at node $q$. From the chain rule with $X=C_q^i\widetilde{X}_q$, we have
\begin{equation}\label{eq:local_lagrange}
  C_q^{iT}\nabla_X\mathcal{L}(C_q^{i}\widetilde{X}_q^{i+1},\widetilde{\bm{\lambda}}(q))=0.
\end{equation}
The KKT conditions for optimality of the local problem (\ref{eq:loc_prob_g_compact}) can then be written as
\begin{align}
  &C_q^{iT}\nabla_X\Big[f\left(C_q^i\widetilde{X}_q^{i+1}\right)+\sum_{j\in\mathcal{J}}\lambda_j(q)h_j\left(C_q^i\widetilde{X}_q^{i+1}\right)\Big]=0, \label{eq:stat_d}\\
  &h_j\left(C_q^i\widetilde{X}_q^{i+1}\right)\leq 0\;\textrm{ $\forall j\in\mathcal{J}_I$, } h_j\left(C_q^i\widetilde{X}_q^{i+1}\right)=0\;\textrm{ $\forall j\in\mathcal{J}_E$,}  \label{eq:pf_d}\\
  &\lambda_j(q)\geq 0\;\textrm{ $\forall j\in\mathcal{J}_I$,} \label{eq:dual_d}\\
  &\lambda_j(q)h_j\left(C_q^i\widetilde{X}_q^{i+1}\right)=0\;\textrm{ $\forall j\in\mathcal{J}_I$,} \label{eq:cs_d}
\end{align}
where the $\lambda_j(q)$'s are the Lagrange multipliers at updating node $q$ and iteration $i$. Since (\ref{eq:pf_d}) is exactly the same as (\ref{eq:pf}), we conclude that the local primal feasibility condition is also satisfied globally (which we already knew from (\ref{eq:loc_glob}) and \cite[Lemma 1]{musluoglu2022unifiedp1}). Let us now look at the three other equations and assume that the algorithm has reached a fixed point, i.e., $X^{i+1}=X^i=\widebar{X}$. From this fixed point assumption, we can replace $C_q^i\widetilde{X}_q^{i+1}=X^{i+1}$ with $C_q^i\widetilde{X}_q^i=X^i=\widebar{X}$ in (\ref{eq:stat_d}) such that the local stationarity condition can be rewritten as
\begin{equation}\label{eq:stat_equi}
  C_q^{iT}\nabla_X\Big[f(\widebar{X})+\sum_{j\in\mathcal{J}}\lambda_j(q)h_j(\widebar{X})\Big]=0.
\end{equation}
Selecting the first $M_q$ rows of (\ref{eq:stat_equi}), we have (note that from (\ref{eq:cqi_tree}), the matrix $C_q^{iT}$ selects the $q-$th block-row from $X$ as the first $M_q$ rows)
\begin{equation}\label{eq:Aq_stat}
  \nabla_{X_q}f(\widebar{X})=-\sum_{j\in\mathcal{J}}\lambda_j(q)\nabla_{X_q}h_j(\widebar{X}).
\end{equation} 
Since this result is valid for any node $q$ due to the fixed point assumption, we may stack the variations of equation (\ref{eq:Aq_stat}) for every node $q\in\mathcal{K}$:
\begin{equation}\label{eq:stat_mat}
  \resizebox{.45\textwidth}{!}{%
  $\begin{bmatrix}
    \nabla_{X_1}f(\widebar{X})\\
    \vdots\\
    \nabla_{X_K}f(\widebar{X})
  \end{bmatrix}=\nabla_Xf(\widebar{X})=-\begin{bmatrix}
    \sum_{j\in\mathcal{J}}\lambda_j(1)\nabla_{X_1}h_j(\widebar{X})\\
    \vdots\\
    \sum_{j\in\mathcal{J}}\lambda_j(K)\nabla_{X_K}h_j(\widebar{X})
  \end{bmatrix}.$%
  }
\end{equation}
Multiplying (\ref{eq:stat_equi}) by $\widetilde{X}_q^i$ defined in (\ref{eq:X_fixed}) and using the fact that $C_q^i\widetilde{X}_q^i=X^i=\widebar{X}$ (this follows from (\ref{eq:X_fixed}) and the definition of $C_q^i$ in (\ref{eq:cqi_tree})-(\ref{eq:Theta_qi_def})), we obtain
\begin{equation}\label{eq:lin_indep_1}
  \widebar{X}^{T}\nabla_X f(\widebar{X})=-\sum_{j\in\mathcal{J}}\lambda_j(q)\widebar{X}^{T}\nabla_X h_j(\widebar{X}).
\end{equation}
From Condition \ref{cond:lin_indep_1}, the set $\{\widebar{X}^{T}\nabla_X h_j(\widebar{X})\}_j$ is linearly independent and therefore the Lagrange multipliers $\{\lambda_j(q)\}_j$ that satisfy (\ref{eq:lin_indep_1}) are unique. Moreover, since the left-hand side of (\ref{eq:lin_indep_1}) does not depend on the node $q$, we have that $\lambda_j(q)=\lambda_j$ for any node $q$. Therefore, (\ref{eq:stat_mat}) becomes
\begin{equation}\label{eq:glob_stat_equi}
  \nabla_X f(\widebar{X})=-\sum_{j\in\mathcal{J}}\lambda_j \nabla_X h_j(\widebar{X}),
\end{equation}
which implies that $(\widebar{X},\{\lambda_j\}_j)$ satisfy the global stationarity conditions (\ref{eq:stat}). Since $(\widebar{X},\{\lambda_j\}_j)$ satisfies the local dual feasibility and the local complementary slackness conditions (\ref{eq:dual_d}) and (\ref{eq:cs_d}) (with $C_q^i\widetilde{X}_q^{i+1}=X^{i+1}$ replaced by $\widebar{X}$ due to it being a fixed point), it also satisfies their global counterparts (\ref{eq:dual}) and (\ref{eq:comp_slack}). Hence, $(\widebar{X},\{\lambda_j\}_j)$ satisfies all the global KKT optimality conditions. This proves that any fixed point of Algorithm \ref{alg:dasf} is a stationary point of the global problem (\ref{eq:prob_g_compact}).

\end{proof}

\section{ }\label{app:dasf_conv_2}

\begin{proof}[Proof of Theorem \ref{thm:stationarity_2}]
The arguments in this proof are very similar to the ones in the proof of Theorem \ref{thm:stationarity}, where the main difference is to show the uniqueness of the Lagrange multipliers when $\{D_{j,q}(\widebar{X})\}_j$ is a linearly independent set for any $q\in\mathcal{K}$ at fixed points $\widebar{X}=X^{i+1}=X^i$. Therefore we only make changes to that part of the proof.

From the definition of $C_q^i$ in (\ref{eq:cqi_tree})-(\ref{eq:Theta_qi_def}), left-multiplying each block-row of (\ref{eq:stat_equi}) by the corresponding block-column of $\widetilde{X}_q^{(i+1)T}$ as structured in (\ref{eq:X_tilde_part}), we have
\begin{align}
  X_q^{(i+1)T}\nabla_{X_q} \Big[f(X^i)+\sum_{j\in\mathcal{J}}\lambda_j(q)h_j(X^i)\Big]&=0,\label{eq:node_q_blk}\\
  G_n^{(i+1)T}\sum_{k\in\mathcal{B}_{nq}}X_k^{iT}\nabla_{X_k} \Big[f(X^i)+\sum_{j\in\mathcal{J}}\lambda_j(q)h_j(X^i)\Big]&=0,\label{eq:node_n_blk}
\end{align}
$\forall n\in\mathcal{N}_q$. Note that we here again assume that the algorithm has reached a fixed point for which $X^{i+1}=X^i=\widebar{X}$ (as this was also assumed to derive (\ref{eq:stat_equi})), which implies that we can make the substitutions $X_q^{i+1}=X_q^i=\widebar{X}_q$ and $G_n^{i+1}=I_Q$ for all $n\in\mathcal{N}_q$ (see (\ref{eq:upd_X})) within (\ref{eq:node_q_blk})-(\ref{eq:node_n_blk}). By doing so and, from the definition (\ref{eq:Dqj}) of $D_{j,q}(\widebar{X})$, (\ref{eq:node_q_blk}) and (\ref{eq:node_n_blk}) become 
\begin{equation}\label{eq:stat_D_blk}
  C_{X_q}^{iT}\nabla_{X}f(\widebar{X})+\sum_{j\in\mathcal{J}}\lambda_j(q)D_{j,q}(\widebar{X})=0,
\end{equation}
where $C_{X_q}^i$ is the matrix $C_q^i$ but the identity matrix of the first block-column in (\ref{eq:cqi_tree}) has been replaced by $X_q^{i}=\widebar{X}_q$. From (\ref{eq:stat_D_blk}) and the linear independence assumption over the set $\{D_{j,q}(\widebar{X})\}_j$ (see Condition \ref{cond:lin_indep_2}), the Lagrange multipliers $\lambda_j(q)$ for all $j\in\mathcal{J}$ that satisfy (\ref{eq:stat_D_blk}) must be unique. We can repeat the same argument for any updating node, which implies that (\ref{eq:node_q_blk})-(\ref{eq:stat_D_blk}) holds for any node $q$, each time with its own unique set of Lagrange multipliers. We will now prove that this unique set of Lagrange multipliers is the same for any updating node $q$.

Slightly rewriting (\ref{eq:node_q_blk})-(\ref{eq:node_n_blk}) (with $X^{i+1}=X^i=\widebar{X}$ and $G_n^{i+1}=I_Q$) gives
\begin{align}
    \widebar{X}_q^{T}\nabla_{X_q}f(\widebar{X})&=-\sum_{j\in\mathcal{J}}\lambda_j(q)\widebar{X}_q^{T}\nabla_{X_q}h_j(\widebar{X}),\label{eq:stat_qf_fixed}\\
    \sum_{k\in\mathcal{B}_{nq}}\widebar{X}_k^{T}\nabla_{X_k}f(\widebar{X})&=-\sum_{j\in\mathcal{J}}\sum_{k\in\mathcal{B}_{nq}}\lambda_j(q)\widebar{X}_k^{T}\nabla_{X_k}h_j(\widebar{X}),\label{eq:stat_qh_fixed}
\end{align}
where (\ref{eq:stat_qf_fixed})-(\ref{eq:stat_qh_fixed}) holds for every $q\in\mathcal{K}$ and for all $n\in\mathcal{N}_q$. Substituting (\ref{eq:stat_qf_fixed}) into (\ref{eq:stat_qh_fixed}), we obtain
\begin{equation}
\begin{aligned}
    &\sum_{j\in\mathcal{J}}\sum_{k\in\mathcal{B}_{nq}}\lambda_j(k)\widebar{X}_k^{T}\nabla_{X_k}h_j(\widebar{X})\\=&\sum_{j\in\mathcal{J}}\sum_{k\in\mathcal{B}_{nq}}\lambda_j(q)\widebar{X}_k^{T}\nabla_{X_k}h_j(\widebar{X}).\label{eq:lin_system_h}
\end{aligned}
\end{equation}
Vectorizing the matrices in (\ref{eq:lin_system_h}) such that $\mathbf{h}_{j,k}=\text{vec}\left(\widebar{X}_k^{T}\nabla_{X_k}h_j(\widebar{X})\right)\in\mathbb{R}^{Q^2}$ and defining $H_k=[\mathbf{h}_{1,k},\dots,\mathbf{h}_{J,k}]\in\mathbb{R}^{Q^2\times J}$ $\forall k\in\mathcal{K}$, we obtain
\begin{equation}
    \sum_{k\in\mathcal{B}_{nq}}H_k\bm{\lambda}(k)=\left(\sum_{k\in\mathcal{B}_{nq}}H_k\right)\bm{\lambda}(q),
\end{equation}
where $\bm{\lambda}(k)=[\lambda_1(k),\dots,\lambda_J(k)]^T$. At node $q$ and for its corresponding neighbor $n\in\mathcal{N}_q$, we can then write the following linear system of equation:
\begin{equation}
    \mathbf{H}_{nq}\cdot\bm{\lambda}_{\mathcal{K}}=0,
\end{equation}
where $\bm{\lambda}_{\mathcal{K}}=[\bm{\lambda}^T(1),\dots,\bm{\lambda}^T(K)]^T\in\mathbb{R}^{KJ}$ and $\mathbf{H}_{nq}\in\mathbb{R}^{Q^2\times KJ}$ is a block-column matrix where the $l-$th block of size $Q^2\times J$ is given by
\begin{equation}\label{eq:H_nq_blocks}
    \mathbf{H}_{nq}(l)=\begin{cases}
    -\sum_{k\in\mathcal{B}_{nq}}H_k & \text{if $l=q$}\\
    H_l & \text{if $l\in\mathcal{B}_{nq}$}\\
    0 & \text{otherwise.}
    \end{cases}
\end{equation}
Stacking vertically the matrices $\mathbf{H}_{nq}$, for every neighbor $n\in\mathcal{N}_q$ and every node $q\in\mathcal{K}$, results in $\mathbf{H}\in\mathbb{R}^{Q^2\sum_{k}|\mathcal{N}_k|\times KJ}$ and we write
\begin{equation}\label{eq:H_lambda_0}
    \mathbf{H}\cdot \bm{\lambda}_{\mathcal{K}}=0.
\end{equation}
Note that from (\ref{eq:H_nq_blocks}), the sum of all $Q^2\times J$ block-columns of $\mathbf{H}_{nq}(l)$ must be equal to the zero matrix. Therefore, every $\bm{\lambda}_{\mathcal{K}}$ such that $\bm{\lambda}(1)=\dots=\bm{\lambda}(K)$ is in the null space of $\mathbf{H}$ and would satisfy (\ref{eq:H_lambda_0}). The dimension of the set $\{\bm{\lambda}_{\mathcal{K}}\in\mathbb{R}^{KJ}\;|\; \bm{\lambda}(1)=\dots=\bm{\lambda}(K)\}$ is equal to $J$ and therefore $\text{rank}(\mathbf{H})\leq KJ-J$. To ensure that these are the only solutions of (\ref{eq:H_lambda_0}), we require $\text{rank}(\mathbf{H})=KJ-J$. Note that a necessary condition to satisfy this is that $KJ-J\leq Q^2\sum_{k}|\mathcal{N}_k|$, i.e., $KJ-J$ is less than the number of rows of $\mathbf{H}$, or equivalently $J\leq\frac{Q^2}{K-1}\sum_{k}|\mathcal{N}_k|$, leading to the upper bound given in (\ref{eq:J_upper_bound_2_2}).

\begin{lem}\label{lem:rank_H}
If Condition \ref{cond:lin_indep_2} holds, then $\text{rank}(\mathbf{H})=KJ-J$.
\end{lem}
\begin{proof}
The proof of this lemma is provided as supplementary material to the paper as it is too elaborate to fit in this main text.
\end{proof}

Since $\text{rank}(\mathbf{H})=KJ-J$, the set $\{\bm{\lambda}_{\mathcal{K}}\in\mathbb{R}^{KJ}\;|\; \bm{\lambda}(1)=\dots=\bm{\lambda}(K)\}$ contains the full null space of $\mathbf{H}$ and hence all the solutions of (\ref{eq:H_lambda_0}). Because we now have established that all the Lagrange multipliers are the same, we can conclude that (\ref{eq:stat_mat}) holds in this case as well, allowing to obtain the same result as in (\ref{eq:glob_stat_equi}). The remaining arguments of the proof of Theorem \ref{thm:stationarity} can be applied here to conclude that each fixed point is a point satisfying the KKT conditions (\ref{eq:stat})-(\ref{eq:comp_slack}) of the original problem (\ref{eq:prob_g_compact}).
\end{proof}

\section{ }\label{app:continuity}

\begin{proof}[Proof of Theorem \ref{thm:cons_cont}]

    From Corollary \ref{cor:X0_compact}, all points in $(X^i)_i$ remain in a compact set. Since each compact set has at least one accumulation point $\widebar{X}$, there exists an infinite subsequence of $(X^i)_i$ which converges to $\widebar{X}$. Because the number of nodes is finite, there exists a node $k\in\mathcal{K}$ that acts as an updating node in an infinite number of iterations that are sampled in this subsequence. In other words, we can find some node $k$ such that there exists a set of iteration indices $\mathcal{I}_k\subseteq\mathbb{N}$ such that $(X^i)_{i\in\mathcal{I}_k}$ converges to $\widebar{X}$ and $(i\mod K)_{i\in\mathcal{I}_k}=(k)_{i\in\mathcal{I}_k}$, i.e., the iteration indices $\mathcal{I}_k$ correspond only to iterations where node $k$ is the updating node in Algorithm \ref{alg:dasf}.
    From Berge's Maximum Theorem \cite{berge1997topological}, the continuity of $\widetilde{\mathcal{F}}_k$ and $g(W,V)\triangleq ||W-V||_F$ implies that $\widetilde{\mathcal{M}}_k$ and thus $\mathcal{M}_k=C_k\widetilde{\mathcal{M}}_k$ as in (\ref{eq:m_minimum_norm2}) are upper semicontinuous. This implies that any accumulation point $\widebar{X}^{(+1)}$ of the set $(X^{i+1})_{i\in\mathcal{I}_k}$ must satisfy
    \begin{equation}
            \label{eq:limit_in_mq}
            \widebar{X}^{(+1)}\in\mathcal{M}_k(\widebar{X}).
    \end{equation}
    As from Lemma \ref{lem:local_fixed_points} (see end of this appendix), we have $\mathcal{M}_k(\widebar{X})=\{\widebar{X}\}$ (i.e., $\widebar{X}$ is a fixed point of $\mathcal{M}_k$), (\ref{eq:limit_in_mq}) implies that 
    \begin{equation}
            \widebar{X}^{(+1)}=\widebar{X}.
    \end{equation}
    Inductively applying the above argument for the new subsequence $(X^{i+1})_{i\in\mathcal{I}_k}$ and accumulation point $\widebar{X}^{(+1)}$ and for node $k+1 \mod K$ yields 
    \begin{equation}
            \widebar{X}^{(+l)}=\widebar{X}\quad\forall l\geq0.
    \end{equation}
    As $\widebar{X}^{(+l)}$ is \textit{any} accumulation point $(X^{i+l})_{i\in\mathcal{I}_k}$, all the accumulation points of $(X^{i+l})_{i\in\mathcal{I}_k}$ are equal to $\widebar{X}$ and all the sequences 
    \begin{equation}
            (X^{i})_{i\in\mathcal{I}_{(k+l\mod K)}}\triangleq(X^{i+l})_{i\in\mathcal{I}_k}
    \end{equation}
    converge to the same point $\widebar{X}$. From Lemma \ref{lem:local_fixed_points} (here applied to the node $k+l \mod K$ instead of $k$), $\widebar{X}^{(+l)}$ is a fixed point of $\mathcal{M}_{(q+l\mod K)}$ and $\widebar{X}$ is therefore a fixed point of $\mathcal{M}_k$ for any $k$, proving the first part of the theorem. 

    We now prove that $\lim_{i\rightarrow+\infty}||X^{i+1}-X^i||_F=0$ by contradiction. 
    Let us assume that the above statement is not true. We first note that $ X,W\to ||X-W||_F$ is a continuous mapping and $X^i$ and $X^{i+1}$ both live in a compact set (see beginning of the proof). Since the continuous image of a compact set is itself a compact set, $(||X^{i+1}-X^{i}||_F)_{i}$ has at least one accumulation point. 
    There must therefore be some index set $\mathcal{I}$ such that
    \begin{equation}\label{eq:diff_x_non_zero}
            \lim_{i\in\mathcal{I}\to\infty} ||X^{i+1}-X^{i}||_F > 0,
    \end{equation}
    that is, there is one convergent subsequence converging to a point different from zero. Indeed, if zero was the only accumulation point, the sequence would be convergent (see Lemma \ref{lem:conv_to_a}).
    Furthermore, based on the same reasoning as the beginning of this proof, there is some $\mathcal{I}_k'\subseteq \mathcal{I}$ such that $(X^i)_{i\in\mathcal{I}_k'}$ is a convergent sequence such that $(i\mod K)_{i\in\mathcal{I}_k'}=(k)_{i\in\mathcal{I}_k'}$ for some $k$. We have shown above that the convergence of such a sequence $(X^i)_{i\in\mathcal{I}_k'}$ implied that
    \begin{align}\label{eq:same_limit}
            \lim_{i\in\mathcal{I}_k'\to\infty} X^i = \lim_{i\in\mathcal{I}_k'\to\infty} X^{i+1}.
    \end{align}
    Therefore, by continuity of the Frobenius norm, it must be that
    \begin{equation}\label{eq:diff_to_0}
            \lim_{i\in\mathcal{I}_k'\to\infty} ||X^{i+1}-X^{i}||_F = 0.
    \end{equation}
    As (\ref{eq:diff_to_0}) contradicts (\ref{eq:diff_x_non_zero}), every convergent subsequence of $(||X^{i+1}-X^{i}||_F)_{i\in\mathbb{N}}$ converges to $0$ and $(||X^{i+1}-X^{i}||_F)_{i\in\mathbb{N}}$ is therefore a convergent sequence.

\end{proof}

\begin{proof}[Proof of Theorem \ref{thm:conv_single_pt}]
    Let us assume that the mapping (\ref{eq:alg}) corresponding to the DASF algorithm has a finite set of fixed points denoted $\Phi$. As the set of fixed points is finite, it must be that there exists some $\delta>0$ such that for any pair of fixed points $\widebar{X},\widebar{W}$ $||\widebar{X}-\widebar{W}||_F>\delta$. From Lemma \ref{lem:conv_to_a}, as the sublevel sets of $f$ are compact, $(X^i)_i$ converges to the set of its accumulation points $\mathcal{A}$. From Theorem \ref{thm:cons_cont}, this set $\mathcal{A}$ must be a subset of $\Phi$, and therefore finite. 
    
    We will now show that $\mathcal{A}$ must be a singleton. From Lemma \ref{lem:conv_to_a}, we have
    \begin{equation}\label{eq:dist_from_set_acc}
        \forall \varepsilon, \exists i_\varepsilon>0:  \inf_{W\in\mathcal{A}}||W-X^i||_F<\varepsilon, \quad \forall i>i_\varepsilon.
    \end{equation}
    From Theorem \ref{thm:cons_cont}, we have
    \begin{equation}\label{eq:dist_from_set_acc2}
        \forall \varepsilon, \exists i_\varepsilon>0: ||X^{i+1}-X^i||_F < \varepsilon, \quad \forall i > i_\varepsilon.
    \end{equation}
    From (\ref{eq:dist_from_set_acc}) and (\ref{eq:dist_from_set_acc2}), there exists an $i_\varepsilon>0$ such that
    \begin{align}
        \forall i>i_\varepsilon\;\exists \widebar{X}, \widebar{X}^{(+1)}\in\mathcal{A}:\nonumber &||\widebar{X}-X^i||_F<\delta/3,\\\nonumber
        &||\widebar{X}^{(+1)}-X^{i+1}||_F<\delta/3,\\
                       & ||X^{i+1}-X^i||_F < \delta/3. \label{eq:cond_deltas}
    \end{align}
    We then have from the triangle inequality
    \begin{align}
    \label{eq:triangle_eq}
        &||\widebar{X}^{(+1)} - \widebar{X}||_F\leq||\widebar{X}^{(+1)}-X^i||_F\nonumber\\ & + \norm{X^{i+1}-X^i}_F +||X^{i+1}-\widebar{X}^{(+1)}||_F < \delta.
    \end{align}
    If $\widebar{X} \neq \widebar{X}^{(+1)}$ then this would imply that $||\widebar{X}^{(+1)} - \widebar{X}||_F>\delta$ (by definition of $\delta$). However, this would be a contradiction with (\ref{eq:triangle_eq}). Therefore, $\widebar{X}$ and $\widebar{X}^{(+1)}$ must be equal. Since (\ref{eq:cond_deltas}) holds for any $i$, we find by induction that $\mathcal{A}$ is a singleton. From Lemma \ref{lem:conv_to_a}, this results in the convergence of $(X^i)_i$.
\end{proof}

\begin{lem}\label{lem:local_fixed_points}
Let $\mathcal{I}_k\subseteq\mathbb{N}$ be such that $(X^i)_{i\in\mathcal{I}_k}$ converges to $\widebar{X}$ and $(i\mod K)_{i\in\mathcal{I}_k}=(k)_{i\in\mathcal{I}_k}$, i.e., we only consider iterates related to some node $k$. 
Then if $\widetilde{\mathcal{F}}_k: \mathbb{R}^{M\times Q}\rightrightarrows \mathbb{R}^{\widetilde{M}_Q\times Q}$ is a continuous mapping\footnote{As seen in the proof of Lemma \ref{lem:local_fixed_points}, this can be relaxed to upper semicontinuity, but we keep the continuity condition for consistency with previous results.}, $\widebar{X}$ is a fixed point of the map $\mathcal{M}_k$, i.e., $\mathcal{M}_k(\widebar{X})=\{\widebar{X}\}$.
\end{lem}
\begin{proof}
    From Corollary \ref{cor:X0_compact}, all points in $(X^i)_i$ remain in a compact set therefore $(X^{i+1})_{i\in\mathcal{I}_k}$ has an accumulation point $\widebar{X}^{(+1)}$. The continuity of $\widetilde{\mathcal{F}}_k$, and thus of $C_k(X)\widetilde{\mathcal{F}}_k$, implies that it is also upper semicontinuous. Therefore we have (by definition, see \cite{charalambos2013infinite})
    \begin{equation}\label{eq:usc_F}
        \widebar{X}^{(+1)}\in C_k(\widebar{X})\widetilde{\mathcal{F}}_k(\widebar{X}).
    \end{equation}
    We can now prove that
    \begin{equation}\label{eq:f_plus_one}
    \min_{W\in \mathcal{S}_k(\widebar{X})}f(W) = f(\widebar{X}^{(+1)})=f(\widebar{X}).
    \end{equation}
    The first equality directly follows from (\ref{eq:usc_F}) and the definition \eqref{def_FQ} of $\widetilde{\mathcal{F}}_k$, that is
    \begin{equation}
        \min_{W\in \mathcal{S}_k(\widebar{X})}f(W)=\min_{\widetilde{W}_q:C_q(\widebar{X})\widetilde{W}_q\in \mathcal{S}_k(\widebar{X})}f(C_q(\widebar{X})\widetilde{W}_q)=f(X)
    \end{equation}
    $\forall X\in C_k(\widebar{X})\widetilde{\mathcal{F}}_k(\widebar{X})$. The second equality in \eqref{eq:f_plus_one} follows from the fact that $\widebar{X}$ is an accumulation point and that $f$ is continuous together with the fact that $(f(X^i))_i$ is monotonically decreasing (Lemma \ref{lem:monotonic}) (i.e., all accumulation points have the same objective value).
    Because of (\ref{eq:f_plus_one}), and since $\widebar{X}$ is by definition in $\mathcal{S}_k(\widebar{X})$, it must be that $\widebar{X}\in C_k(\widebar{X})\widetilde{\mathcal{F}}_k(\widebar{X})$ and thus $[\widebar{X}_k^T,I_Q,\dots,I_Q]^T\in\widetilde{\mathcal{F}}_k(\widebar{X})$. Using this in (\ref{eq:m_minimum_norm}) with $X$ replaced by $\widebar{X}$ results in $\{\widebar{X}\}=\mathcal{M}_k(\widebar{X})$.
\end{proof}

\begin{lem}\label{lem:conv_to_a}
   In a compact metric space, a sequence converges to the set of its accumulation points. Additionally, the sequence converges if and only if it has a single accumulation point.
\end{lem}
\begin{proof}
    Let $(X^i)_i$ be some sequence in a compact metric space. Let $\mathcal{A}$ denote the set of accumulation points of $(X^i)_i$. We have $\forall X^*\in\mathcal{A}, \exists \mathcal{I}\subseteq \mathbb{N}:\forall \varepsilon >0, \exists i_\varepsilon >0: ||X^*-X^i||_F<\varepsilon, \quad \forall i\in\mathcal{I}>i_\varepsilon$.
    We wish to prove that $\forall \varepsilon>0,\exists i_\varepsilon>0 : \inf_{X\in\mathcal{A}}||X-X^i||_F<\varepsilon,\quad \forall i > i_\varepsilon$.
    Let us assume that our claim is not true. Then
    \begin{equation}
        \exists \varepsilon>0, \forall i_\varepsilon>0,\exists i>i_\varepsilon : \inf_{X\in\mathcal{A}}||X-X^i||_F\geq\varepsilon.
    \end{equation}
    which implies that there exists some infinite set $\mathcal{I}\subseteq\mathbb{N}$ such that
    \begin{equation}\label{eq:subsequence_away_from_A}
            \exists i_\varepsilon>0: \inf_{X\in\mathcal{A}}||X-X^i||_F\geq \varepsilon, \quad \forall i\in\mathcal{I}>i_\varepsilon.
    \end{equation}
    Since the space is compact, the subsequence $(X^i)_{i\in{\mathcal{I}}}$ has itself a convergent subsequence converging to a point in $\mathcal{A}$, contradicting (\ref{eq:subsequence_away_from_A}).

    The convergence in the case of a single accumulation point follows directly from the previous result and the converse is a well-known result.
\end{proof}




\bibliographystyle{IEEEtran}
\bibliography{IEEEabrv,IEEEexample}

\end{document}


\setcounter{equation}{98}
\setcounter{figure}{2}
\setlength{\textfloatsep}{5pt plus 1.0pt minus 2.0pt}

\title{A Unified Algorithmic Framework for Distributed Adaptive Signal and Feature Fusion Problems \\\LARGE--- Part II: Convergence Properties: Supplementary Material}

\author{Cem Ates~Musluoglu, Charles Hovine, 
        and~Alexander~Bertrand,~\IEEEmembership{Senior~Member,~IEEE}
\thanks{Copyright \copyright 2023 IEEE. Personal use of this material is permitted. Permission from IEEE must be obtained for all other uses, in any current or future media, including reprinting/republishing this material for advertising or promotional purposes, creating new collective works, for resale or redistribution to servers or lists, or reuse of any copyrighted component of this work in other works.}
\thanks{This paper is a supplementary material to \cite{musluoglu2022unifiedp2}.}
}

\maketitle

This supplementary material details the steps we follow to prove Lemma \ref{main-lem:rank_H} from \cite{musluoglu2022unifiedp2}, i.e., to show that $\text{rank}(\mathbf{H})=KJ-J$ which implies that the equation $\mathbf{H}\cdot \bm{\lambda}_{\mathcal{K}}=0$ (equation (\ref{main-eq:H_lambda_0}) in \cite{musluoglu2022unifiedp2}) can only have solutions $\bm{\lambda}_{\mathcal{K}}=[\bm{\lambda}^T(1),\dots,\bm{\lambda}^T(K)]^T$ of the form $\bm{\lambda}(1)=\dots=\bm{\lambda}(K)$ when Condition \ref{main-cond:lin_indep_2} (repeated below) is satisfied.

\begin{figure}[t]
  \includegraphics[width=0.48\textwidth]{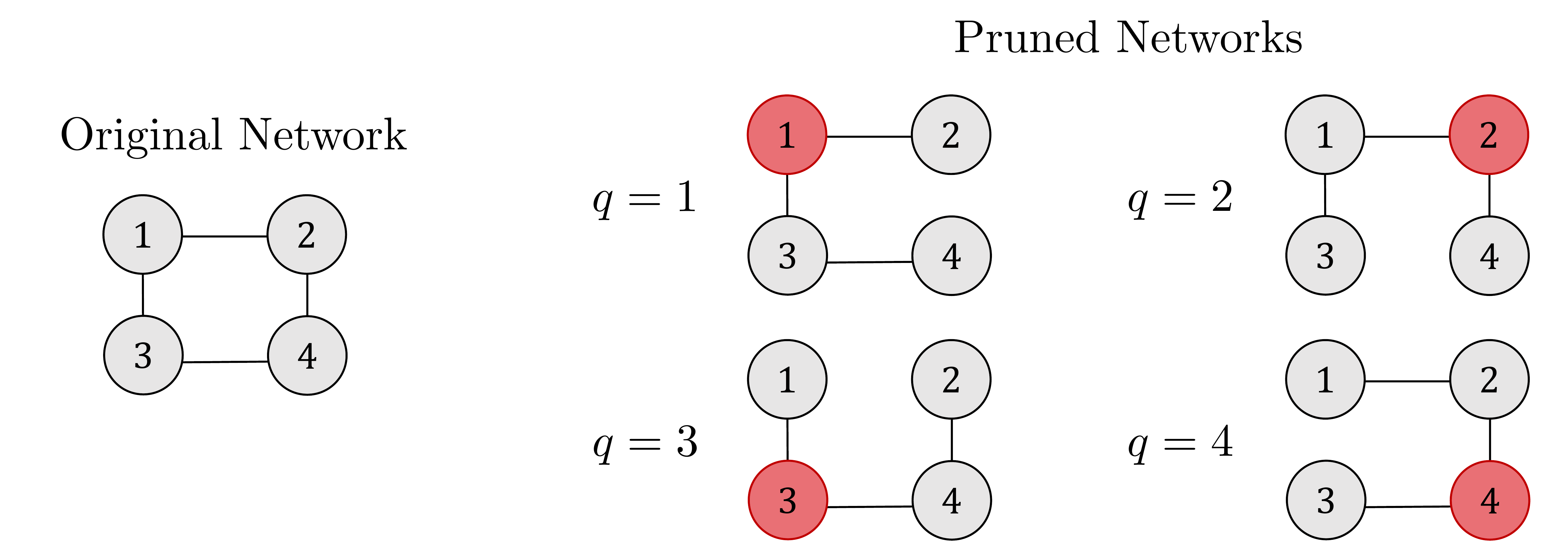}
  \caption{The example $4-$node network we will use to illustrate the equations.}
  \label{fig:example_network}
\end{figure}

\begin{namedcond}[Condition \ref{main-cond:lin_indep_2}]
For a fixed point $\widebar{X}$ of Algorithm \ref{main-alg:dasf}, the elements of the set $\{D_{j,q}(\widebar{X})\}_{j\in\mathcal{J}}$ are linearly independent for any $q$, where
\begin{equation}\label{eq:Dqj}
  D_{j,q}(\widebar{X})=\begin{bmatrix}
    \widebar{X}_q^{T}\nabla_{X_q} h_j(\widebar{X})\\
    \sum\limits_{k\in\mathcal{B}_{n_1q}}\widebar{X}_k^{T}\nabla_{X_k} h_j(\widebar{X})\\
    \vdots\\
    \sum\limits_{k\in\mathcal{B}_{n_{|\mathcal{N}_q|}q}}\widebar{X}_k^{T}\nabla_{X_k} h_j(\widebar{X})
  \end{bmatrix},
\end{equation}
which is a block-matrix containing $(1+|\mathcal{N}_q|)$ blocks of $Q\times Q$ matrices.
\end{namedcond}
The proof will be accompanied by an example network topology, given in Figure \ref{fig:example_network}, to visualize the structure of some large matrices in the proof, yet we keep the proof itself generic.

At a fixed point $\widebar{X}=X^{i+1}=X^i$ of the DASF algorithm, we have shown in Appendix \ref{main-app:dasf_conv_2} in \cite{musluoglu2022unifiedp2} that at each node $q\in\mathcal{K}$ the local stationarity conditions can be written as
\begin{align}
    \widebar{X}_q^{T}\nabla_{X_q}f(\widebar{X})&=-\sum_{j\in\mathcal{J}}\lambda_j(q)\widebar{X}_q^{T}\nabla_{X_q}h_j(\widebar{X}),\label{eq:stat_qf_fixed}\\
    \sum_{k\in\mathcal{B}_{nq}}\widebar{X}_k^{T}\nabla_{X_k}f(\widebar{X})&=-\sum_{j\in\mathcal{J}}\sum_{k\in\mathcal{B}_{nq}}\lambda_j(q)\widebar{X}_k^{T}\nabla_{X_k}h_j(\widebar{X}),\label{eq:stat_qh_fixed}
\end{align}
$\forall n\in\mathcal{N}_q$, leading to the equations
\begin{equation}
\begin{aligned}
    &\sum_{j\in\mathcal{J}}\sum_{k\in\mathcal{B}_{nq}}\lambda_j(k)\widebar{X}_k^{T}\nabla_{X_k}h_j(\widebar{X})\\=&\sum_{j\in\mathcal{J}}\sum_{k\in\mathcal{B}_{nq}}\lambda_j(q)\widebar{X}_k^{T}\nabla_{X_k}h_j(\widebar{X}),\; \forall n\in\mathcal{N}_q.\label{eq:lin_system_h}
\end{aligned}
\end{equation}

For clarity, we vectorize the the matrices $\widebar{X}_k^{T}\nabla_{X_k}h_j(\widebar{X})$ such that $\mathbf{h}_{j,k}=\text{vec}\left(\widebar{X}_k^{T}\nabla_{X_k}h_j(\widebar{X})\right)\in\mathbb{R}^{Q^2}$ and create the matrix $H_k=[\mathbf{h}_{1,k},\dots,\mathbf{h}_{J,k}]\in\mathbb{R}^{Q^2\times J}$ $\forall k\in\mathcal{K}$. Note that the linear independence condition of $\{D_{j,q}\}_j$ for any node $q$ is then equivalent to
\begin{equation}\label{eq:matrix_D_q}
    \mathbf{D_q}=\begin{bmatrix}
    H_q\\
    \sum_{k\in\mathcal{B}_{n_1q}}H_k\\
    \vdots\\
    \sum_{k\in\mathcal{B}_{n_{|\mathcal{N}_q|}q}}H_k
    \end{bmatrix}
\end{equation}
being full rank, i.e., $\text{rank}(\mathbf{D_q})=J$. Then, (\ref{eq:lin_system_h}) can be rewritten as
\begin{equation}\label{eq:local_equation_lambda}
    \sum_{k\in\mathcal{B}_{nq}}H_k\bm{\lambda}(k)=\left(\sum_{k\in\mathcal{B}_{nq}}H_k\right)\bm{\lambda}(q),\; \forall n\in\mathcal{N}_q,
\end{equation}
where $\bm{\lambda}(k)=[\lambda_1(k),\dots,\lambda_J(k)]^T$. For example, in the example network from Figure \ref{fig:example_network}, we have for $q=1$:
\begin{align}
    &H_2\bm{\lambda}(2)=H_2\bm{\lambda}(1),\\
    &H_3\bm{\lambda}(3)+H_4\bm{\lambda}(4)=(H_3+H_4)\bm{\lambda}(1).
\end{align}

Equation (\ref{eq:local_equation_lambda}) corresponds to the linear system given as
\begin{equation}
    \mathbf{H}_{nq}\cdot\bm{\lambda}_{\mathcal{K}}=0,
\end{equation}
with $\mathbf{H}_{nq}$ a block-row matrix, where each block corresponds to a node $l\in\mathcal{K}$:
\begin{equation}\label{eq:H_nq_blocks}
    \mathbf{H}_{nq}(l)=\begin{cases}
    -\sum_{k\in\mathcal{B}_{nq}}H_k & \text{if $l=q$}\\
    H_l & \text{if $l\in\mathcal{B}_{nq}$}\\
    0 & \text{otherwise.}
    \end{cases}
\end{equation}
We can then stack vertically every $\mathbf{H}_{nq}$ for every neighbor $n\in\mathcal{N}_q$ of $q$, and for every node $q$ to obtain
\begin{equation}\label{eq:H_lambda_0}
    \mathbf{H}\cdot \bm{\lambda}_{\mathcal{K}}=0.
\end{equation}
In the example network of Figure \ref{fig:example_network}, we have the matrix given in (\ref{eq:H_example}), where we separated by horizontal lines the blocks corresponding to each different node for clarity.
\begin{figure*}[htbp]
\begin{equation}\label{eq:H_example}
    \mathbf{H}=\begin{lbmatrix}{1}
    \mathbf{H}_{2,1}\\
    \mathbf{H}_{3,1}\\
    \hline
    \mathbf{H}_{1,2}\\
    \mathbf{H}_{4,2}\\
    \hline
    \mathbf{H}_{1,3}\\
    \mathbf{H}_{4,3}\\
    \hline
    \mathbf{H}_{2,4}\\
    \mathbf{H}_{3,4}
    \end{lbmatrix}=\begin{lbmatrix}{8}
     -H_2 & \vline & H_2 & \vline & 0 & \vline & 0\\
     -H_3-H_4 & \vline & 0 & \vline & H_3 & \vline & H_4\\
     \hline
     H_1 & \vline & -H_1-H_3 & \vline & H_3 & \vline & 0\\
     0 & \vline & -H_4 & \vline & 0 & \vline & H_4\\
     \hline
     H_1 & \vline & 0 & \vline & -H_1 & \vline & 0\\
     0 & \vline & H_2 & \vline & -H_2-H_4 & \vline & H_4\\
     \hline
     H_1 & \vline & H_2 & \vline & 0 & \vline & -H_1-H_2\\
     0 & \vline & 0 & \vline & H_3 & \vline & -H_3
    \end{lbmatrix}
\end{equation}
\hrule
\end{figure*}

Note that $\mathbf{H}$ is a $Q^2\sum_{k\in\mathcal{K}}|\mathcal{N}_k|\times KJ$ matrix and since $\forall q\in\mathcal{K}$, $n\in\mathcal{N}_q$,
\begin{equation}
    \sum_{k\in\mathcal{K}\backslash \{l\}} \mathbf{H}_{nq}(k)=-\mathbf{H}_{nq}(l)
\end{equation}
$\forall l\in\mathcal{K}$, all $\bm{\lambda}_{\mathcal{K}}\in\mathbb{R}^{KJ}$ such that $\bm{\lambda}(1)=\dots=\bm{\lambda}(K)$ is a solution of (\ref{eq:H_lambda_0}). A basis for this solution space is given by
\begin{equation}
    \mathcal{E}=\left\{\begin{bmatrix}
    \mathbf{e}_1\\
    \vdots\\
    \mathbf{e}_1
    \end{bmatrix},\begin{bmatrix}
    \mathbf{e}_2\\
    \vdots\\
    \mathbf{e}_2
    \end{bmatrix},\dots\begin{bmatrix}
    \mathbf{e}_J\\
    \vdots\\
    \mathbf{e}_J
    \end{bmatrix}\right\}\subset \mathbb{R}^{KJ},
\end{equation}
where $\mathbf{e}_k$'s represent the standard basis for $\mathbb{R}^J$. Since $\mathcal{E}$ spans a $J-$dimensional space, we have
\begin{equation}\label{eq:rank_H_KJ_J}
    \text{rank}(\mathbf{H})\leq KJ-J.
\end{equation}
To show that the solution of (\ref{eq:H_lambda_0}) all satisfy $\bm{\lambda}(1)=\dots=\bm{\lambda}(K)$, we need to show that all solutions are in $\text{span}(\mathcal{E})$, i.e., $\text{rank}(\mathbf{H})=KJ-J$, which is stated in Lemma \ref{main-lem:rank_H} from \cite{musluoglu2022unifiedp2}, repeated below and followed by a proof. By the dimensions of $\mathbf{H}$, this is only possible if we have $Q^2\sum_{k\in\mathcal{K}}|\mathcal{N}_k|\geq KJ-J$ or equivalently $J\leq \frac{Q^2}{K-1}\sum_{k\in\mathcal{K}}|\mathcal{N}_k|$.

\begin{namedlem}[Lemma \ref{main-lem:rank_H}]
If Condition \ref{main-cond:lin_indep_2} holds, then $\text{rank}(\mathbf{H})=KJ-J$.
\end{namedlem}

\begin{proof}
Let us take the bottom part of the matrix $\mathbf{H}$ corresponding to the neighbors of the last node $K$, and take the sum over its block-rows $\mathbf{H}_{nK}$, $n\in\mathcal{N}_K$, defined in (\ref{eq:H_nq_blocks}), which results in a new summed block row
\begin{equation}
    \mathbf{H}_{\Sigma K}(l)=\begin{cases}
    -\sum_{k\in\mathcal{K}\backslash\{K\}}H_k & \text{if $l=K$}\\
    H_l & \text{if $l\neq K$}.
    \end{cases}
\end{equation}
In the example network, we have $K=4$, therefore
\begin{equation}
    \mathbf{H}_{\Sigma 4}=\begin{lbmatrix}{8}
    H_1 & \vline & H_2 & \vline & H_3 & \vline & -H_1-H_2-H_3
    \end{lbmatrix}.
\end{equation}
\begin{figure*}[htbp]
\begin{equation}\label{eq:H_tilde_example}
    \widetilde{\mathbf{H}}=\begin{lbmatrix}{1}
    \mathbf{H}_{2,1}\\
    \mathbf{H}_{3,1}\\
    \mathbf{H}_{\Sigma 4}\\
    \hline
    \mathbf{H}_{1,2}\\
    \mathbf{H}_{4,2}\\
    \mathbf{H}_{\Sigma 4}\\
    \hline
    \mathbf{H}_{1,3}\\
    \mathbf{H}_{4,3}\\
    \mathbf{H}_{\Sigma 4}
    \end{lbmatrix}=\begin{lbmatrix}{8}
     -H_2 & \vline & H_2 & \vline & 0 & \vline & 0\\
     -H_3-H_4 & \vline & 0 & \vline & H_3 & \vline & H_4\\
     H_1 & \vline & H_2 & \vline & H_3 & \vline & -H_1-H_2-H_3\\
     \hline
     H_1 & \vline & -H_1-H_3 & \vline & H_3 & \vline & 0\\
     0 & \vline & -H_4 & \vline & 0 & \vline & H_4\\
     H_1 & \vline & H_2 & \vline & H_3 & \vline & -H_1-H_2-H_3\\
     \hline
     H_1 & \vline & 0 & \vline & -H_1 & \vline & 0\\
     0 & \vline & H_2 & \vline & -H_2-H_4 & \vline & H_4\\
     H_1 & \vline & H_2 & \vline & H_3 & \vline & -H_1-H_2-H_3
    \end{lbmatrix}
\end{equation}
\hrule
\end{figure*}
We then insert vertically the matrices $\mathbf{H}_{\Sigma K}$ after each block $\mathbf{H}_{n_{|\mathcal{N}_q|}q}$, $q\neq K$, of $\mathbf{H}$. Note that this insertion cannot change the rank of the matrix, as all the inserted rows are sums of existing rows in $\mathbf{H}$. We also remove the block-rows $[\mathbf{H}^T_{n_1K},\dots,\mathbf{H}^T_{n_{|\mathcal{N}_K|}K}]^T$, i.e., the block-rows corresponding to node $K$, to obtain a new matrix $\widetilde{\mathbf{H}}$. For the example in Figure \ref{fig:example_network}, this results in the matrix given in (\ref{eq:H_tilde_example}). Note that we have
\begin{equation}\label{eq:rank_H_H_tilde}
    \text{rank}(\mathbf{H})\geq \text{rank}(\widetilde{\mathbf{H}})
\end{equation}
since the removal of rows can only reduce the rank.

We will first look at the rank of $\widetilde{\mathbf{H}}$ and derive from it the rank of $\mathbf{H}$. For this, we will apply the Gaussian elimination method using elementary row operations (EROs) which are known to not change the rank. Referring to the block decomposition of example (\ref{eq:H_tilde_example}), $\widetilde{\mathbf{H}}$ has $K$ block-columns, and each of these block-columns of $\widetilde{\mathbf{H}}$ will be referred to as \textbf{the block-column at position} $k\in\mathcal{K}$. We refer to the matrix $[\mathbf{H}^T_{n_1q},\dots,\mathbf{H}^T_{n_{|\mathcal{N}_q|q}},\mathbf{H}_{\Sigma K}]^T$, $q\neq K$, and the resulting matrices obtained by applying EROs to it as \textbf{the submatrix corresponding to node} $q$. For example, in (\ref{eq:H_tilde_example}) the block-column at position $3$ of the submatrix corresponding to node $2$ is equal to $[H_3^T,0,H_3^T]^T$.

For each $q\neq K$, let us sum all block-rows $\mathbf{H}_{nq}$, $n\in\mathcal{N}_q$. The result is then subtracted from the block-row $\mathbf{H}_{\Sigma K}$ leading to the final block-row of the submatrix corresponding to each node $q$ being of the form $[0|\dots|0|\sum_{k\in\mathcal{K}}H_k|0|\dots|0|-\sum_{k\in\mathcal{K}}H_k]$, where the first non-zero matrix is at position $q\neq K$. For our example network, we obtain
\begin{equation}\label{eq:H_tilde_sums}
\scalebox{0.92}{$
    \begin{lbmatrix}{8}
     -H_2 & \vline & H_2 & \vline & 0 & \vline & 0\\
     -H_3-H_4 & \vline & 0 & \vline & H_3 & \vline & H_4\\
     \sum_{k=1}^4 H_k & \vline & 0 & \vline & 0 & \vline & -\sum_{k=1}^4 H_k\\
     \hline
     H_1 & \vline & -H_1-H_3 & \vline & H_3 & \vline & 0\\
     0 & \vline & -H_4 & \vline & 0 & \vline & H_4\\
     0 & \vline & \sum_{k=1}^4 H_k & \vline & 0 & \vline & -\sum_{k=1}^4 H_k\\
     \hline
     H_1 & \vline & 0 & \vline & -H_1 & \vline & 0\\
     0 & \vline & H_2 & \vline & -H_2-H_4 & \vline & H_4\\
     0 & \vline & 0 & \vline & \sum_{k=1}^4 H_k & \vline & -\sum_{k=1}^4 H_k
    \end{lbmatrix}.$}
\end{equation}
An important observation is that, for each $q\neq K$, the block-column at position $q$ corresponding to the submatrix of node $q$ can be obtained by applying EROs to $\mathbf{D_q}$ defined in (\ref{eq:matrix_D_q}), i.e., it is equal to $\mathbf{D_q}$ up to EROs. Since $\text{rank}(\mathbf{D_q})=J$ by Condition \ref{main-cond:lin_indep_2}, we can apply the necessary EROs to obtain the following reduced echelon form for the submatrix corresponding to node $q=1$:
\begin{equation}\label{eq:first_J_pivots}
    \begin{lbmatrix}{10}
    I_J & \vline & \multirow{2}{*}{$*$} & \vline & \multirow{2}{*}{$\cdots$} & \vline & \multirow{2}{*}{$*$} & \vline & \multirow{2}{*}{$*$} &\\
    0 & \vline &  & \vline & & \vline &  & \vline &  &
    \end{lbmatrix}.
\end{equation}
We can use the $J$ pivots in the first $J$ columns of (\ref{eq:first_J_pivots}) to create zeros at all the entries underneath (in the submatrices corresponding to $q\neq 1$) using EROs. As a result, we obtain 

\begin{equation}\label{eq:echelon_q_1}
    \begin{lbmatrix}{10}
    I_J & \vline & \multirow{2}{*}{$*$} & \vline & \multirow{2}{*}{$\cdots$} & \vline & \multirow{2}{*}{$*$} & \vline & \multirow{2}{*}{$*$} &\\
    0 & \vline &  & \vline & & \vline &  & \vline &  &\\
    \hline
    \multirow{2}{*}{$0$} & \vline & \multirow{2}{*}{$\mathcal{R}\mathbf{D_2}$} & \vline & \multirow{2}{*}{$\cdots$} & \vline & \multirow{2}{*}{$*$} & \vline & \multirow{2}{*}{$*$} &\\
     & \vline &  & \vline & & \vline &  & \vline &  &\\
     \hline
     \multirow{2}{*}{$\vdots$} & \vline & \multirow{2}{*}{$\vdots$} & \vline & \multirow{2}{*}{$\ddots$} & \vline & \multirow{2}{*}{$\vdots$} & \vline & \multirow{2}{*}{$\vdots$} &\\
     & \vline &  & \vline & & \vline &  & \vline &  &\\
     \hline
     \multirow{2}{*}{$0$} & \vline & \multirow{2}{*}{$*$} & \vline & \multirow{2}{*}{$\cdots$} & \vline & \multirow{2}{*}{$\mathcal{R}\mathbf{D_{K-1}}$} & \vline & \multirow{2}{*}{$*$} &\\
     & \vline &  & \vline & & \vline &  & \vline &  &
    \end{lbmatrix},
\end{equation}
where $\mathcal{R}\mathbf{D_k}$ is a matrix equal to $\mathbf{D_k}$ up to EROs. To see why (\ref{eq:echelon_q_1}) holds, i.e., that each block-column at position $k\notin\{1,K\}$ of the submatrix corresponding to node $k$ is indeed equal to $\mathbf{D_k}$ up to EROs, we note that, initially, every row above this block is either full of zeros or a row of $H_k$ (see (\ref{eq:H_tilde_sums}) for a visual example). Therefore, the EROs we applied to the full matrix to create zeros at all entries underneath the $J$ pivots in the block-column corresponding to node $q=1$ do not change the fact that the block-column at position $k$ of the submatrix corresponding to node $k$ is equal to $\mathbf{D_k}$ up to EROs. Since EROs do not change the rank of a matrix, the submatrix $\mathcal{R}\mathbf{D_2}$ should again have rank $J$ and so we can again create $J$ pivots to create zeros underneath.
Repeating this process for $2\leq q\leq K-1$, we obtain
\begin{equation}\label{eq:H_tilde_echelon}
    \begin{lbmatrix}{10}
    I_J & \vline & \multirow{2}{*}{$*$} & \vline & \multirow{2}{*}{$\cdots$} & \vline & \multirow{2}{*}{$*$} & \vline & \multirow{2}{*}{$*$} &\\
    0 & \vline &  & \vline & & \vline &  & \vline &  &\\
    \hline
    \multirow{2}{*}{$0$} & \vline & I_J & \vline & \multirow{2}{*}{$\cdots$} & \vline & \multirow{2}{*}{$*$} & \vline & \multirow{2}{*}{$*$} &\\
     & \vline & 0 & \vline & & \vline &  & \vline &  &\\
     \hline
     \multirow{2}{*}{$\vdots$} & \vline & \multirow{2}{*}{$\vdots$} & \vline & \multirow{2}{*}{$\ddots$} & \vline & \multirow{2}{*}{$\vdots$} & \vline & \multirow{2}{*}{$\vdots$} &\\
     & \vline &  & \vline & & \vline &  & \vline &  &\\
     \hline
     \multirow{2}{*}{$0$} & \vline & \multirow{2}{*}{$0$} & \vline & \multirow{2}{*}{$\cdots$} & \vline & I_J & \vline & \multirow{2}{*}{$*$} &\\
     & \vline &  & \vline & & \vline & 0 & \vline &  &
    \end{lbmatrix}.
\end{equation}
Since there are at least $K-1$ block-columns containing $J$ pivots, $\text{rank}(\widetilde{\mathbf{H}})\geq KJ-J$. We previously established in (\ref{eq:rank_H_H_tilde}) that $\text{rank}(\mathbf{H})\geq \text{rank}(\widetilde{\mathbf{H}})$, hence $\text{rank}(\mathbf{H})\geq KJ-J$. We also already established in (\ref{eq:rank_H_KJ_J}) that $\text{rank}(\mathbf{H})\leq KJ-J$, and therefore it should hold that $\text{rank}(\mathbf{H})=KJ-J$, which is what had to be proven.
\end{proof}

\bibliographystyle{IEEEtran}
\bibliography{IEEEabrv,IEEEexample}
\makeatletter\@input{main_aux_content.tex}\makeatother